\documentstyle[12pt]{article}

\title{Radiowave Method of High Energy Neutrino Detection: \\
calculation of the expected event rate}

\author{A.L.~Provorov and I.M.~Zheleznykh \\
Institute for Nuclear Research of the Russian Academy of Sciences, \\
Moscow, 117312, Russia}

\begin{document}

\maketitle
\begin{abstract}
\noindent The sensitivity of an ice radiowave detector to the anticipated
high energy neutrino fluxes is calculated on the basis of a detailed
threshold analysis and computation of the shower production rate. We show
that diffuse neutrinos from Active Galactic Nuclei could be detectable in
a radio detector of 1 km$^2$ area established in Central Antarctica.
\end{abstract}

\section{Introduction}

About 30 years ago Askar'yan proposed a new method for the detection of
high energy particles by means of the Cherenkov coherent radiowave emission
from the negative charge excess of electromagnetic showers generated in air
or dense media \cite{1}. The charge imbalance of a shower is created by the
Compton scattering of shower photons on atomic electrons, the annihilation
of shower positrons in flight and the knock-on process. The percentage of
negative charge excess amounts to $\sim20\%$ at the shower maximum. The
resulting Cherenkov emission by excess electrons is coherent at wavelengths
larger than the shower lateral dimension, i.e. in the radiowave region. In
spite of the very low frequencies compared with visible light, this
emission should be observable for sufficiently high primary energy because
the radiated power scales with the square of the shower size.

The radio pulses have been indeed successfully observed from air showers
(EAS) in coincidence with particle arrays, but some their properties
(polarization, pulse shape, south-north asymmetry of pulse amplitudes,
etc.) indicate that the radio emission has a different dominant mechanism
of generation in air. Nevertheless, by theoretical reasons the shower
charge excess should be the principal source of the coherent radio
emission in dense media, and after a while interest in the idea of
Askar'yan was renewed by the suggestion to detect high energy (HE)
neutrinos in cold antarctic ice, which has very low radiowave absorption at
temperatures below $-50^\circ$C \cite{2}. It was argued that a radio
antenna array placed on the glacier surface in Central Antarctica could
provide an effective target volume of the order of $10^9-10^{10}$~m$^3$ for
cosmic neutrinos with energies above 100~TeV \cite{2}--\cite{4}.

In this paper we make a detailed threshold analysis for such a detector,
taking account of radiowave absorption in ice. We calculate as well a
production rate of showers initiated by upward-going HE neutrinos in the
upper layer of the Earth. That allows us to estimate an expected neutrino
event rate in a radiowave detector from the different anticipated HE
neutrino fluxes.

\section{Event rate in a radio detector}

We consider a radiowave antarctic neutrino detector as a number of
downward-looking antennae disposed on several dozen meters triangular grid
enclosing a glacier area of the order of 1~km$^2$. The antennae sample
Cherenkov radio pulses from ice, which would provide well-defined conic-type
images on the grid (the Cherenkov angle is equal to 56$^\circ$ in ice)
\cite{2}--\cite{4}.

In general, the radio detector event rate induced by the neutrino flux is
given by

$$
N_\nu=\int dE_o\int d\Omega\,V_{eff}(E_o,\theta,\phi)p_\nu(E_o,\theta,\phi).
\eqno (1)
$$

\noindent Here $p_\nu(E_o,\theta,\phi)$ is the electromagnetic shower
production rate per unit volume for a given shower energy $E_o$ and
direction $(\theta,\phi)$, $V_{eff}(E_o,\theta,\phi)$ is the effective
detection volume.

To make an initial estimation of the event rate we neglect the angular
dependence of $V_{eff}$ and use the next approximation:

$$
V_{eff}=S[R_{max}(E_o)-R_{min}],
\eqno (2)
$$

\noindent where $S$ is the enclosed grid area, $R_{max}(E_o)$ is the
maximum detection depth for a given shower energy $E_o$, $R_{min}$ is the
minimum detection depth, which depends mainly on the grid spacing. In this
case Exp.~(1) is simplified:

$$
N_\nu=\int\limits_{>E_{min}} dE_o\,V_{eff}(E_o)P_\nu(E_o),
\eqno (3)
$$

\noindent where

$$
P_\nu(E_o)=\int d\Omega\,p_\nu(E_o,\theta,\phi)
$$

\noindent and the minimum detection energy $E_{min}$ corresponds to the
minimum detection depth $R_{min}$.

\section{Threshold analysis}

A numerical real time computation of the radiowave emission from
electromagnetic showers developed in the totally transparent ice \cite{5}
gives the following parametrization of the electric field spectrum at the
Cherenkov angle:

$$
R|\vec E(\omega,R,\theta_{\check C})|=\frac{0.55\times10^{-7}
(\nu/\nu_o)}{1+0.4(\nu/\nu_o)^2}\,\frac{E_o}{1\mbox{ TeV}}
\mbox{ (V/MHz)}.
\eqno (4)
$$

\noindent Here $\nu$ is the frequency, $R$ is the distance from the shower,
$E_o$ is the incident electron (photon) energy, $\nu_o=500$ MHz (the result
of \cite{5} was divided by 2 to define the Fourier transform as $\vec
E(\omega)=\int dt\,\vec E(t)exp(i\omega t)$).

Radiowave attenuation in the real ice strongly depends on the wavelength,
as well as on the ice temperature. Therefore, we have used the original
data on radiowave absorption in ice \cite{6} together with the results of
temperature profile measurements in a super deep bore hole at the Vostok
Antarctic Station \cite{7} to calculate the total attenuation of a shower
radio pulse after vertical propagation from a given depth to the ice surface
(Fig.~1).

To obtain the threshold energy for the one-channel radiowave detection of
electromagnetic showers we need to consider in some detail a process of
radio pulse transformations by a receiving antenna and its preamplifier
(active filter). The relation between output antenna voltage $V(\omega)$
and incident electric field $\vec E(\omega)$ is given by

$$
V(\omega)=\vec R_A(\omega)\cdot\vec E(\omega),
\eqno (5)
$$

\noindent where $\vec R_A(\omega)$ is the reception transfer function of
antenna. As it is recognized, the so-called TEM horn is the most promising
broad-band antenna for impulsive field measurements (for instance, see
Ref.~\cite{8}). Its reception transfer function has approximately constant
magnitude and linear phase dependence over the frequency range from a
hundred MHz to several GHz. As one can see (Fig.~1), the shower radio pulse
spectrum has the same band of wavelengths. Therefore, the TEM horn will
produce an output voltage that is a high fidelity replica of the shower
electric field in the time domain. For example, the $1\times1$ m$^2$
arcsine TEM horn specially designed for the neutrino radiowave experiments
has $R_A\simeq0.14$ V/(V/m) for a normal incident field in air. The
reception pattern of this antenna is rather broad with the half-amplitude
beamwidths of about $90^\circ$ in both $E$ and $H$ planes for an incident
electromagnetic pulse of 1~ns duration \cite{4}.

The conventional definition of a signal-to-noise ratio at the filter output
is \cite{9}

$$
\frac{S}{N}=\frac{\mbox{peak instantaneous output signal power}}
{\mbox{output noise power}}.
\eqno (6)
$$

\noindent The maximum of (6) occurs when the filter transfer function is
proportional to the complex conjugation of the input filter voltage
$V^*(\omega)$ (``matched filter''). The maximum value is

$$
[S/N]_{max}=\frac{2}{N_o}\int\limits_{-\infty}^{+\infty}
\frac{d\omega}{2\pi}|V(\omega)|^2,
\eqno (7)
$$

\noindent where $N_o$ is the one-sided white noise power density at the
filter input. From (5) and (7) we obtain for the normal incident shower
radio pulse received by a TEM horn in ice:

$$
[S/N]_{max}=\frac{2}{N_o}\varepsilon R_A^2
\int\limits_{-\infty}^{+\infty} \frac{d\omega}{2\pi}|\vec
E(\omega,R,\theta_{\check C})|^2,
\eqno (8)
$$

\noindent where $\varepsilon$ is the relative permeability of ice ($R_A$
rises by a factor of $\sqrt{\varepsilon}$ in medium). If the antenna
impedance is equal to the load (filter) one,

$$
N_o=kT_NZ_L.
\eqno (9)
$$

\noindent Here $k=1.381023$~J/K is the Boltzmann constant, $T_N$ is the
noise temperature and $Z_L$ is the load impedance. Using the
parametrization (4) and taking into account the radiowave absorption in ice
for vertical pulse propagation from 100~m depth, we find for $Z_L=50$~Ohm
and $\sqrt{\varepsilon}=1.8$ (ice refraction coefficient):

$$
[S/N]_{max}\simeq0.1\frac{E_o^2(\mbox{TeV})}{T_N(\mbox{K})}
\left(\frac{R_A}{0.14\mbox{ m}}\right)^2.
\eqno (10)
$$

\noindent For the signal-to-noise ratio of unity, $T_N=300$~K (according to
the antarctic noise measurements \cite{10}) and $R_A=0.14$~V/(V/m) the
shower threshold energy $E_{th}$ is approximately equal to 55~TeV. This is
a factor of 7 lower than the result obtained by Zas, Halzen and Stanev
\cite{5}. They performed the threshold estimation for a half-wave dipole
antenna adapted for the narrow-band receiving technique of EAS
radio detection experiments. Hence, the use of a broad-band TEM horn
together with a matched filter result in the significant decrease of the
threshold energy.

Fig.~2 shows the calculated dependence of the relative threshold energy on
the shower production depth. The related results of the maximum detection
depth $R_{max}$ can be parameterized by the expression:

$$
R_{max}(E_o)\simeq615\ln(1+E_o/5.8E_{100})\mbox{ (m)},
\eqno (11)
$$

\noindent where $E_{100}$ is the threshold energy for the depth of 100~m.

\section{Production rate of neutrino-induced showers}

Electromagnetic showers are directly produced only by the charged current
interactions of electron neutrinos:

$$
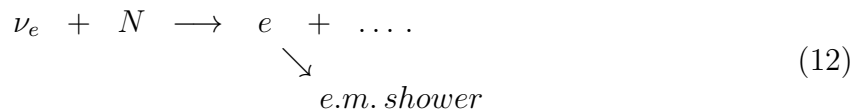
\begin{array}{ccccccc}
\nu_e& +& N& \longrightarrow& e& +& \ldots\, . \\
&&&& \quad \lefteqn{\searrow} \\
&&&&& \lefteqn{e.m.\, shower}
\end{array}
\eqno (12)
$$

\noindent But hadronic showers, which are generated by all the types of
neutrino interactions

$$
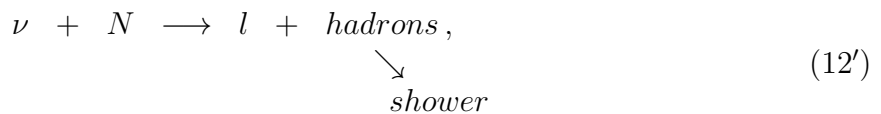
\begin{array}{ccccccc}
\nu& +& N& \longrightarrow& l& +& hadrons\, , \\
&&&&&& \searrow \\
&&&&&& \lefteqn{shower}
\end{array}
\eqno (12')
$$

\noindent initiate electromagnetic subshowers too through decay of neutral
pions and eta particles. According to the accelerator calorimetric
experiments and Monte Carlo simulations, the contribution of
electromagnetic subshowers to the total energy deposit of hadronic shower
increases with the initial hadron energy and amounts to $\sim90\%$ at
$E_o>10$ TeV \cite{11}. As the mean inelasticity in the reaction ($12'$) is
$\sim0.3$ for $E_\nu>100$ Tev \cite{12}, one should expect that the power
in the radio signal from the neutrino-induced hadronic shower will be only
slightly less than the power from the pure electromagnetic shower of the
same energy. In the paper we neglect this difference, though the question
deserves more detailed study.

The shower production rate $P_\nu(E,\gamma)$ for upward-going neutrinos is
defined as

$$
P_\nu(E_o,\gamma)=n_N\sigma_o G_\nu^{tot}(E_o,\gamma)\Phi(E_o,\gamma).
\eqno (13)
$$

\noindent Here $G_\nu^{tot}(E_o,\gamma)$ is the total shower production
function, $\Phi(E_o,\gamma)$ is the differential power-law energy spectrum
of incident neutrino flux with the integral spectral index $\gamma$, $n_N$
is the nucleon number per unit volume; $\sigma_o=1.1\times10^{-34}$~cm$^2$
is the standard cross section.

The total shower production functions $G_\nu^{tot}(E_o,\gamma)$ for
electron and muon neutrino fluxes are given by

$$
G_{\nu_e}^{tot}(E_o,\gamma)=\frac{\sigma^{cc}(E_o)}{\sigma_o}
\Omega_{eff}(E_o,\gamma)+G_\nu^{nc}(E_o,\gamma)
\eqno (14)
$$

$$
G_{\nu_\mu}^{tot}(E_o,\gamma)=G_\nu^{cc}(E_o,\gamma)+G_\nu^{nc}(E_o,\gamma)
\eqno (14')
$$

\noindent respectively with the definition of $G_\nu^j(E_o,\gamma)$
($j\equiv cc$ for the charged current interaction and $j\equiv nc$ for the
neutral one) as

$$
G_\nu^j(E_o,\gamma)=\int\limits_0^1dy\,y^\gamma
\Omega_{eff}(E_o/y,\gamma)\frac{\partial\sigma^j(E_o/y,y)}
{\sigma_o\partial y}
\eqno (15)
$$

\noindent In these expressions $\sigma^j(E)$ is the neutrino-nucleon cross
section, $y=(E_\nu-E_l)/E_\nu$ is the relative energy loss of neutrino in
the laboratory frame and $\Omega_{eff}(E,\gamma)$ is the effective solid
angle for upward-going neutrino flux, which is defined through the
coefficient of neutrino absorption by the Earth $k(E,\theta,\gamma)$ as
in~\cite{13}

$$
\Omega_{eff}(E,\gamma)=2\pi\int\limits_0^1d\cos\theta\,k(E,\theta,\gamma).
\eqno (16)
$$

\noindent Let us notice that the shower production functions
$G_\nu^j(E_o,\gamma)$ are actually the generalization of the neutrino
spectrum-weighted moments

$$
Z_\nu^j(E_o,\gamma)=\int\limits_0^1dy\,y^\gamma
\frac{\partial\sigma^j(E_o/y,y)}{\sigma_o\partial y},
\eqno (17)
$$

\noindent which have been originally calculated in the paper~\cite{14},
for the case of neutrino absorption.  We performed the computations of
$G_\nu^{tot}(E_o,\gamma)$ by use of the EHQL~\cite{15} and DLA~\cite{16}
parametrizations of the quark distribution functions, as it has been
proposed for the neutrino cross section evaluation in \cite{17,12}, and
PREM model of the Earth structure \cite{18}. The results for electron and
muon neutrino fluxes with $\gamma=2.1$ are given in Fig.~3 and Fig.~4
respectively.

All the functions $G_\nu^j(E_o,\gamma)$ vary rather slowly with $\gamma$.
The behavior of the $\bar\nu_e$ shower production function around 6.4~PeV
(Fig.~4) is strongly affected by the Glashow resonance reaction

$$
\bar\nu_e+e^-\to W^-\to all
\eqno (19)
$$

\noindent with a very high cross section \cite{19,20}. The Earth is almost
opaque for such neutrinos.

\section{Detection of AGN neutrinos}

Active Galactic Nuclei (AGN) have long ago been considered as possible
cosmic objects for high energy neutrino production \cite{21,22}. Due to the
recent development of the idea in some important ways by Stecker et al.
\cite{23,24} and other authors \cite{25}--\cite{27}, nowadays AGN are
recognized as most promising sources of HE neutrinos. Therefore, we
calculate the expected event rate in a 1~km$^2$ radiowave detector for the
different existing models of AGN neutrino production, assuming
$E_{100}=55$~TeV and $R_{min}=100$~m (for comparative discussion of the
models see Ref.~\cite{28,29}). The results are given in Table~1. The
expected background from atmospheric neutrinos \cite{30} will be $\sim20$
events per year.

\vspace{2mm}
\noindent Table~1. Expected event rate for the different models of AGN
neutrino production
\begin{center}
\begin{tabular}{|l||c|}
\hline
\bf Diffuse neutrino fluxes &
\bf Expected event rates ($\bf S/N=1$), \\
\bf from AGN by &
\bf year$\bf^{-1}$ km$\bf^{-2}$ \\
\hline
Stecker et al., 1992 & 110 \\
\hline
Szabo and Protheroe, 1992 (max) & 2800 \\
\hline
Szabo and Protheroe, 1992 (min) & 730 \\
\hline
Biermann, 1992 & 110 \\
\hline
Sikora and Begelman, 1992 & 100 \\
\hline
\end{tabular}
\end{center}
\vspace{2mm}

These figures can be compared with the predicted muon event rate in a
10$^4$~m$^2$ optical underwater neutrino telescope, such as DUMAND~II,
NESTOR or NT-200, which are under construction now. According to \cite{29}
at muon energies above 1~TeV the rate will be for Ref.~\cite{24} $\sim30$
per year and for Ref.~\cite{25} from 160 to 800, decreasing by several
times at 10~TeV. Ultra high energy neutrinos from AGN can be detected by
DUMAND~II due to the Glashow resonance reaction with the rate for
Ref.~\cite{24} about several dozen events per year \cite{31}.

Fig.~5 shows a plot of rate in a radiowave detector versus signal-to-noise
ratio. As one can see, even at high detection thresholds the diffuse AGN
flux remains measurable by a radiowave detector for all the models, save
the Sikora and Begelman one.

We have estimated the event rate in a radiowave detector only for
upward-going neutrino fluxes, though inclined neutrinos from the upper
hemisphere would give rise to the event rate too, if some part of initiated
Cherenkov radio emission has a direction toward the ice surface.  However,
these calculations are very sensitive to the particular form of the angular
dependence of the effective detection volume and require a full Monte Carlo
simulation of the detector response.

Besides of the surface deployment of a radio array, it was proposed also to
put antennae deep under the ice in bore holes to detect neutrinos from the
upper hemisphere more effectively \cite{32}. Because of the limiting hole
size ($20''$ diameter), only small conical antennae are suitable in this
case and, therefore, the detection threshold should be several times
higher. However, a downward-going HE neutrino flux does not suffer from the
Earth shadowing effect and this can partially compensate the event rate
decrease, especially due to the Glashow resonance reaction. Some type of a
combined detector is possible also.

\section{Conclusions}

In this paper we have shown that a 1~km$^2$ radiowave neutrino detector
established in Central Antarctica (at the South Pole or Vostok Station)
should be sensitive to the predicted diffuse fluxes of AGN neutrinos at
energies above a hundred TeV. For some production models AGN neutrinos
would be effectively detected in the broad energy region up to 1~PeV, that
gives, in principle, a possibility to determine the spectrum shape.

The performed calculations of the shower production rate by upward-going HE
neutrino flux are relevant also for the other HE neutrino detection
experiments.

\section*{Acknowledgments}

The authors wish to thank F.~Halzen, D.W.~McKay, R.K.~Moore, J.P.~Ralston
and E.~Zas for helpful discussions. We also would like to thank
V.S.~Berezinsky and L.V.~Volkova for supplying data on HE neutrino fluxes.
We are very grateful to M.A.~Markov for his continual encouragement and our
colleagues I.N.~Boldyrev and V.O.~Pozovnoy for the many useful remarks.

\clearpage

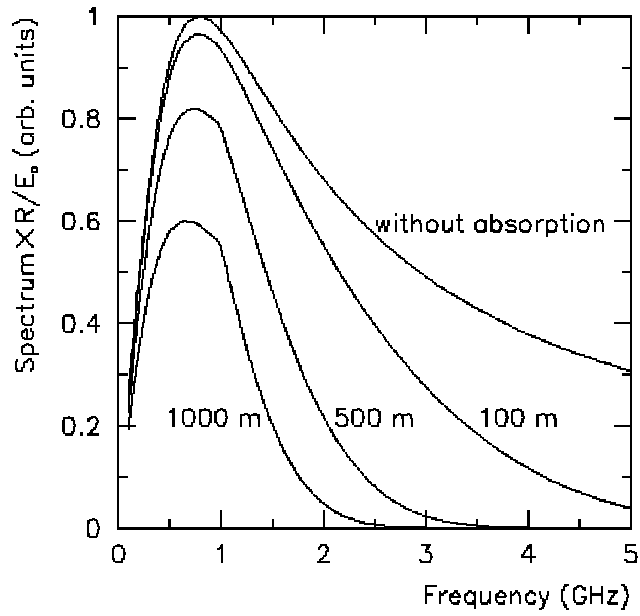
\begin{figure}[t]
\begin{center}
\begin{picture}(800,800)(0,0)\ifx\higzdraft\undefined\newcount\higzdraft\higzdraft=0{}\fi\ifnum\higzdraft>0
\put(0,0){\framebox(800,800){}}\else\ifx\higzstep\undefined\newcount\higzstep\higzstep=0{}\fi\ifnum\higzstep<1\higzstep=2
\fi\ifx\higzxx\undefined\newcount\higzxx\newcount\higzyy\newcount\higzx\newcount\higzy\newcount\higzdx\newcount\higzdy
\newcount\higzlx\newcount\higzly\newcount\higzslope\newcount\higzlen\newcount\higzllen\newcount\higzoffs\newcount\higzloffs
\newcount\higzadash\newcount\higzbdash\newcount\higzcdash\newcount\higzddash\newcount\higzmsize\newcount\higztemp\fi
\def\higzstroke#1,#2,#3,#4;{\advance\higzloffs\higzllen\ifnum\higzloffs>#1\advance\higzloffs-\higzllen\advance\higzloffs-#1
\higzloffs=-\higzloffs\ifnum#2>0\put(\higzlx,\higzly){\line(#3,#4){\higzloffs}}\fi\ifnum#2<0\put(\higzlx,\higzly){\circle*{0}}\fi
\higztemp=\higzloffs\multiply\higztemp#3\advance\higzlx\higztemp\higztemp=\higzloffs\multiply\higztemp#4\advance\higzly\higztemp
\advance\higzllen-\higzloffs\higzloffs=#1\else\ifnum#2>0\put(\higzlx,\higzly){\line(#3,#4){\higzllen}}\fi\ifnum#2<0\put(
\higzlx,\higzly){\circle*{0}}\fi\higzllen=0\fi}\def\higzdashed#1,#2,#3,#4,#5;{{\higzlx=#1\higzly=#2\higzllen=#5\higzloffs=
\higzoffs\loop\ifnum\higzloffs<\higzadash\ifnum\higzadash>1\higzstroke\higzadash,1,#3,#4;\else\higzstroke\higzadash,-1,#3,#4;\fi
\else\ifnum\higzloffs<\higzbdash\higzstroke\higzbdash,0,#3,#4;\else\ifnum\higzloffs<\higzcdash\higztemp=\higzcdash\advance
\higztemp-\higzbdash\ifnum\higztemp>1\higzstroke\higzcdash,1,#3,#4;\else\higzstroke\higzcdash,-1,#3,#4;\fi\else\ifnum\higzloffs<
\higzddash\higzstroke\higzddash,0,#3,#4;\else\higzloffs=0\fi\fi\fi\fi\ifnum\higzllen>0\repeat\global\higzoffs=\higzloffs}}\def
\higzsolid#1,#2,#3,#4,#5;{\put(#1,#2){\line(#3,#4){#5}}}\def\higzhslant#1,#2,#3;{\higzslope=#1\multiply\higzslope1000\advance
\higzslope500\divide\higzslope#2\higzlen=\higzslope\multiply\higzlen\higzstep\divide\higzlen1000\higzdy=0\loop\ifnum
\higzdy<#2\higzx=\higzxx\higzy=\higzyy\higzdx=\higzslope\multiply\higzdx\higzdy\advance\higzdx500\divide\higzdx1000\advance
\higzy\higzdy\multiply\higzdx#3\advance\higzx\higzdx\multiply\higzdx#3\advance\higzdx\higzlen\ifnum\higzdx>#1\advance
\higzlen#1\advance\higzlen-\higzdx\fi\higzline\higzx,\higzy,#3,0,\higzlen;\advance\higzdy\higzstep\repeat}\def\higzvslant#1,#2,#3;{
\higzslope=#2\multiply\higzslope1000\advance\higzslope500\divide\higzslope#1\higzlen=\higzslope\multiply\higzlen\higzstep
\divide\higzlen1000\higzdx=0\loop\ifnum\higzdx<#1\higzx=\higzxx\higzy=\higzyy\higzdy=\higzslope\multiply\higzdy\higzdx\advance
\higzdy500\divide\higzdy1000\advance\higzx\higzdx\multiply\higzdy#3\advance\higzy\higzdy\multiply\higzdy#3\advance
\higzdy\higzlen\ifnum\higzdy>#2\advance\higzlen#2\advance\higzlen-\higzdy\fi\higzline\higzx,\higzy,0,#3,\higzlen;\advance\higzdx
\higzstep\repeat}\def\s#1,#2;{\higzdx=#1{}\ifnum\higzdx<0\higzdx=-\higzdx\fi\higzdy=#2{}\ifnum\higzdy<0\higzdy=-\higzdy\fi\ifnum
\higzdx<\higzdy\ifnum#1<0\advance\higzxx#1\advance\higzyy#2\ifnum#2<0\higzvslant-#1,-#2,1;\else\higzvslant-#1,#2,-1;
\fi\else\ifnum#2<0\higzvslant#1,-#2,-1;\else\higzvslant#1,#2,1;\fi\advance\higzxx#1\advance\higzyy#2\fi\else\ifnum#2<0
\advance\higzxx#1\advance\higzyy#2\ifnum#1<0\higzhslant-#1,-#2,1;\else\higzhslant#1,-#2,-1;\fi\else\ifnum#1<0
\higzhslant-#1,#2,-1;\else\higzhslant#1,#2,1;\fi\advance\higzxx#1\advance\higzyy#2\fi\fi}\def\h#1;{\higzline\higzxx,
\higzyy,1,0,#1;\advance\higzxx#1}\def\r#1;{\higzline\higzxx,\higzyy,-1,0,#1;\advance\higzxx-#1}\def\U#1;{\higzline\higzxx,
\higzyy,0,1,#1;\advance\higzyy#1}\def\D#1;{\higzline\higzxx,\higzyy,0,-1,#1;\advance\higzyy-#1}\def\m#1,#2;{\higzxx=#1
\higzyy=#2}\def\higzdot#1,#2;{\put(#1,#2){\circle*{\higzmsize}}}\def\higzplus#1,#2;{\higzx=#1\multiply\higzx2\advance\higzx-
\higzmsize\divide\higzx2\put(\higzx,#2){\line(1,0){\higzmsize}}\higzy=#2\multiply\higzy2\advance\higzy-\higzmsize\divide
\higzy2\put(#1,\higzy){\line(0,1){\higzmsize}}}\def\higzstar#1,#2;{\higzplus#1,#2;\higzcross#1,#2;}\def
\higzcircle#1,#2;{\put(#1,#2){\circle{\higzmsize}}}\def\higzcross#1,#2;{\let\higzsave\higzline\let\higzline\higzsolid\higzlx=#1
\multiply\higzlx2\advance\higzlx-\higzmsize\divide\higzlx2\higzly=#2\multiply\higzly2\advance\higzly-\higzmsize\divide
\higzly2\m\higzlx,\higzly;\s\higzmsize,\higzmsize;\higzly=#2\multiply\higzly2\advance\higzly\higzmsize\divide\higzly2\m\higzlx,
\higzly;\s\higzmsize,-\higzmsize;\let\higzline\higzsave}\def\p#1,#2;{\higzmarker#1,#2;}\def\f#1,#2;{\put(\higzxx,
\higzyy){\makebox(#1,#2)[lb]{\rule{#1\unitlength}{#2\unitlength}}}}
\let\higzline=\higzsolid\m579,93;\D23;\m579,93;\h14;\m579,82;\h8;\m598,86;\D16;\m598,79;\s1,3;\s3,2;\s2,2;\h3;\m611,79;\h13;\U2;
\s-1,2;\s-3,3;\r3;\s-2,-2;\s-3,-2;\s-1,-3;\D2;\s1,-3;\s3,-3;\s2,-1;\h3;\s2,1;\s2,3;\m644,86;\D23;\m644,82;\s-2,2;\s-3,2;\r3;\s-4,-4;
\s-1,-3;\D2;\s1,-3;\s2,-3;\s2,-1;\h3;\s3,1;\s2,3;\m652,86;\D11;\s1,-4;\s3,-1;\h3;\s2,1;\s3,4;\m664,86;\D16;\m672,79;\h13;\U2;\s-1,2;
\s-1,1;\s-3,2;\r3;\s-4,-4;\s-1,-3;\D2;\s1,-3;\s2,-3;\s2,-1;\h3;\s3,1;\s2,3;\m692,86;\D16;\m692,81;\s4,3;\s2,2;\h3;\s2,-2;\s1,-3;
\D11;\m725,82;\s-2,2;\s-3,2;\r3;\s-4,-4;\s-1,-3;\D2;\s1,-3;\s2,-3;\s2,-1;\h3;\s3,1;\s2,3;\m730,86;\s7,-16;\m743,86;\s-6,-16;\s-2,-4;
\s-3,-2;\s-2,-1;\r1;\m768,97;\s-2,-2;\s-2,-3;\s-2,-4;\s-2,-6;\D4;\s4,-10;\s2,-3;\s2,-2;\m791,88;\s-1,2;\s-3,2;\s-2,1;\r4;\s-2,-1;
\s-2,-2;\s-1,-2;\s-1,-4;\D5;\s1,-3;\s1,-2;\s2,-3;\s2,-1;\h4;\s2,1;\s3,3;\s1,2;\U3;\m785,79;\h6;\m798,93;\D23;\m813,93;\D23;\m798,82;
\h15;\m833,86;\s-12,-16;\m821,86;\h12;\m821,70;\h12;\m839,97;\s3,-2;\s2,-3;\s2,-4;\s1,-6;\D4;\s-1,-5;\s-2,-5;\s-2,-3;\s-3,-2;
\m37,395;\s-2,-3;\s-1,-3;\D4;\s1,-3;\s2,-3;\h2;\s3,1;\s2,4;\s2,6;\s1,2;\s3,3;\h4;\s2,-3;\s1,-3;\D4;\s-1,-3;\s-2,-3;\m42,402;\h22;
\m45,402;\s-2,2;\s-1,2;\U4;\s1,2;\s2,2;\s3,1;\h2;\s4,-1;\s2,-2;\s1,-2;\D4;\s-1,-2;\s-2,-2;\m48,422;\U13;\r2;\s-3,-3;\s-1,-2;\D3;
\s1,-2;\s2,-2;\s3,-1;\h2;\s4,1;\s2,2;\s1,2;\U3;\s-1,2;\s-2,3;\m45,454;\s-2,-2;\s-1,-2;\D4;\s1,-2;\s2,-2;\s3,-1;\h2;\s4,1;\s2,2;
\s1,2;\U4;\s-1,2;\s-2,2;\m34,463;\h18;\s4,1;\s1,2;\U2;\m42,459;\U8;\m42,475;\h15;\m48,475;\s-3,1;\s-2,2;\s-1,2;\U3;\m42,489;\h10;
\s4,1;\s1,2;\U3;\s-1,2;\s-4,3;\m42,500;\h15;\m42,509;\h15;\m46,509;\s-3,3;\s-1,3;\U3;\s1,2;\s3,1;\h11;\m46,521;\s-3,3;\s-1,2;\U4;
\s1,2;\s3,1;\h11;\m34,545;\s23,23;\m34,568;\s23,-23;\m34,579;\h23;\m34,579;\U10;\s1,3;\s3,3;\h3;\s3,-3;\s1,-3;\D10;\m45,587;\s12,8;
\m30,619;\s34,-19;\m34,626;\h23;\m34,626;\U14;\m45,626;\U9;\m57,626;\U14;\m55,646;\D1;\s1,-1;\h5;\s1,1;\U4;\s-1,1;\s-2,1;\r1;
\s-2,-1;\s-1,-1;\D3;\m30,675;\s2,-2;\s3,-2;\s4,-2;\s6,-2;\h4;\s6,2;\s4,2;\s3,2;\s2,2;\m42,695;\h15;\m45,695;\s-2,-3;\s-1,-2;\D3;
\s1,-2;\s2,-2;\s3,-1;\h2;\s4,1;\s2,2;\s1,2;\U3;\s-1,2;\s-2,3;\m42,703;\h15;\m48,703;\s-3,1;\s-2,2;\s-1,3;\U3;\m34,717;\h23;\m45,717;
\s-2,2;\s-1,3;\U3;\s1,2;\s2,2;\s3,1;\h2;\s4,-1;\s2,-2;\s1,-2;\D3;\s-1,-3;\s-2,-2;\m55,739;\s1,-1;\s1,1;\s-1,1;\s-1,-1;\m42,759;\h10;
\s4,1;\s1,3;\U3;\s-1,2;\s-4,3;\m42,771;\h15;\m42,780;\h15;\m46,780;\s-3,3;\s-1,2;\U4;\s1,2;\s3,1;\h11;\m34,799;\s1,1;\s-1,2;\s-1,-2;
\s1,-1;\m42,800;\h15;\m34,810;\h18;\s4,1;\s1,2;\U3;\m42,807;\U7;\m45,833;\s-2,-1;\s-1,-3;\D4;\s1,-3;\s2,-1;\s2,1;\s1,2;\s1,6;\s1,2;
\s2,1;\h2;\s2,-1;\s1,-3;\D4;\s-1,-3;\s-2,-1;\m30,839;\s2,3;\s3,2;\s4,2;\s6,1;\h4;\s6,-1;\s4,-2;\s3,-2;\s2,-3;\m514,580;\s4,-15;
\m523,580;\s-5,-15;\m523,580;\s4,-15;\m531,580;\s-4,-15;\m538,588;\s1,-1;\s1,1;\s-1,1;\s-1,-1;\m539,580;\D15;\m549,588;\D18;\s1,-4;
\s2,-1;\h2;\m545,580;\h8;\m560,588;\D23;\m560,576;\s4,3;\s2,1;\h3;\s2,-1;\s1,-3;\D11;\m585,580;\s-2,-1;\s-2,-2;\s-1,-3;\D2;\s1,-4;
\s2,-2;\s2,-1;\h4;\s2,1;\s2,2;\s1,4;\U2;\s-1,3;\s-2,2;\s-2,1;\r4;\m602,580;\D10;\s1,-4;\s2,-1;\h3;\s2,1;\s3,4;\m613,580;\D15;
\m623,588;\D18;\s1,-4;\s2,-1;\h3;\m620,580;\h7;\m658,580;\D15;\m658,577;\s-2,2;\s-3,1;\r3;\s-2,-1;\s-2,-2;\s-1,-3;\D2;\s1,-4;\s2,-2;
\s2,-1;\h3;\s3,1;\s2,2;\m666,588;\D23;\m666,577;\s3,2;\s2,1;\h3;\s2,-1;\s2,-2;\s1,-3;\D2;\s-1,-4;\s-2,-2;\s-2,-1;\r3;\s-2,1;\s-3,2;
\m698,577;\s-1,2;\s-4,1;\r3;\s-3,-1;\s-1,-2;\s1,-2;\s2,-1;\s6,-1;\s2,-1;\s1,-2;\D2;\s-1,-2;\s-4,-1;\r3;\s-3,1;\s-1,2;\m710,580;
\s-3,-1;\s-2,-2;\s-1,-3;\D2;\s1,-4;\s2,-2;\s3,-1;\h3;\s2,1;\s2,2;\s1,4;\U2;\s-1,3;\s-2,2;\s-2,1;\r3;\m726,580;\D15;\m726,574;\s1,3;
\s2,2;\s2,1;\h4;\m740,580;\D22;\m740,577;\s2,2;\s2,1;\h3;\s3,-1;\s2,-2;\s1,-3;\D2;\s-1,-4;\s-2,-2;\s-3,-1;\r3;\s-2,1;\s-2,2;
\m762,588;\D18;\s1,-4;\s2,-1;\h2;\m758,580;\h8;\m772,588;\s1,-1;\s2,1;\s-2,1;\s-1,-1;\m773,580;\D15;\m786,580;\s-2,-1;\s-2,-2;
\s-1,-3;\D2;\s1,-4;\s2,-2;\s2,-1;\h4;\s2,1;\s2,2;\s1,4;\U2;\s-1,3;\s-2,2;\s-2,1;\r4;\m803,580;\D15;\m803,576;\s3,3;\s2,1;\h3;\s2,-1;
\s1,-3;\D11;\m238,325;\s2,1;\s3,3;\D23;\m263,329;\s-4,-1;\s-2,-3;\s-1,-6;\D3;\s1,-5;\s2,-3;\s4,-2;\h2;\s3,2;\s2,3;\s1,5;\U3;\s-1,6;
\s-2,3;\s-3,1;\r2;\m284,329;\s-3,-1;\s-2,-3;\s-1,-6;\D3;\s1,-5;\s2,-3;\s3,-2;\h2;\s4,2;\s2,3;\s1,5;\U3;\s-1,6;\s-2,3;\s-4,1;\r2;
\m306,329;\s-3,-1;\s-3,-3;\s-1,-6;\D3;\s1,-5;\s3,-3;\s3,-2;\h2;\s3,2;\s2,3;\s1,5;\U3;\s-1,6;\s-2,3;\s-3,1;\r2;\m333,322;\D16;
\m333,317;\s3,4;\s2,1;\h3;\s3,-1;\s1,-4;\D11;\m345,317;\s3,4;\s2,1;\h3;\s3,-1;\s1,-4;\D11;\m472,329;\r10;\s-1,-10;\s1,2;\s3,1;\h3;
\s3,-1;\s3,-3;\s1,-3;\D2;\s-1,-3;\s-6,-4;\r3;\s-3,2;\s-1,1;\s-1,2;\m488,329;\s-4,-1;\s-2,-3;\s-1,-6;\D3;\s1,-5;\s2,-3;\s4,-2;\h2;
\s3,2;\s2,3;\s1,5;\U3;\s-1,6;\s-2,3;\s-3,1;\r2;\m509,329;\s-3,-1;\s-2,-3;\s-1,-6;\D3;\s1,-5;\s2,-3;\s3,-2;\h2;\s4,2;\s2,3;\s1,5;\U3;
\s-1,6;\s-2,3;\s-4,1;\r2;\m536,322;\D16;\m536,317;\s4,4;\s2,1;\h3;\s2,-1;\s1,-4;\D11;\m548,317;\s3,4;\s3,1;\h3;\s2,-1;\s1,-4;\D11;
\m653,325;\s3,1;\s3,3;\D23;\m678,329;\s-3,-1;\s-2,-3;\s-1,-6;\D3;\s1,-5;\s2,-3;\s3,-2;\h2;\s4,2;\s2,3;\s1,5;\U3;\s-1,6;\s-2,3;
\s-4,1;\r2;\m700,329;\s-3,-1;\s-2,-3;\s-2,-6;\D3;\s2,-5;\s2,-3;\s3,-2;\h2;\s3,2;\s2,3;\s2,5;\U3;\s-2,6;\s-2,3;\s-3,1;\r2;\m727,322;
\D16;\m727,317;\s3,4;\s2,1;\h4;\s2,-1;\s1,-4;\D11;\m739,317;\s3,4;\s2,1;\h3;\s3,-1;\s1,-4;\D11;\put(170,170){\framebox(681,681){}}
\m170,170;\U681;\m193,170;\r23;\m182,238;\r12;\m193,306;\r23;\m182,375;\r12;\m193,443;\r23;\m182,511;\r12;\m193,579;\r23;\m182,647;
\r12;\m193,715;\r23;\m182,783;\r12;\m193,851;\r23;\m136,182;\s-4,-1;\s-2,-4;\s-1,-5;\D3;\s1,-6;\s2,-3;\s4,-1;\h2;\s3,1;\s2,3;\s1,6;
\U3;\s-1,5;\s-2,4;\s-3,1;\r2;\m103,318;\s-3,-1;\s-2,-4;\s-1,-5;\D3;\s1,-6;\s2,-3;\s3,-1;\h2;\s4,1;\s2,3;\s1,6;\U3;\s-1,5;\s-2,4;
\s-4,1;\r2;\m121,297;\s-2,-1;\s2,-1;\s1,1;\s-1,1;\m130,312;\U1;\s1,3;\s1,1;\s3,1;\h4;\s2,-1;\s1,-1;\s1,-3;\D2;\s-1,-2;\s-2,-3;
\s-11,-11;\h15;\m103,454;\s-3,-1;\s-2,-3;\s-1,-6;\D3;\s1,-5;\s2,-4;\s3,-1;\h2;\s4,1;\s2,4;\s1,5;\U3;\s-1,6;\s-2,3;\s-4,1;\r2;
\m121,433;\s-2,-1;\s2,-1;\s1,1;\s-1,1;\m140,454;\s-11,-15;\h16;\m140,454;\D23;\m103,590;\s-3,-1;\s-2,-3;\s-1,-6;\D3;\s1,-5;\s2,-3;
\s3,-1;\h2;\s4,1;\s2,3;\s1,5;\U3;\s-1,6;\s-2,3;\s-4,1;\r2;\m121,570;\s-2,-1;\s2,-1;\s1,1;\s-1,1;\m143,587;\s-1,2;\s-3,1;\r2;\s-4,-1;
\s-2,-3;\s-1,-6;\D5;\s1,-4;\s2,-2;\s4,-1;\h1;\s3,1;\s2,2;\s1,3;\U1;\s-1,3;\s-2,2;\s-3,2;\r1;\s-4,-2;\s-2,-2;\s-1,-3;\m103,726;
\s-3,-1;\s-2,-3;\s-1,-5;\D4;\s1,-5;\s2,-3;\s3,-1;\h2;\s4,1;\s2,3;\s1,5;\U4;\s-1,5;\s-2,3;\s-4,1;\r2;\m121,706;\s-2,-1;\s2,-1;\s1,1;
\s-1,1;\m135,726;\s-4,-1;\s-1,-2;\D2;\s1,-2;\s2,-1;\s5,-1;\s3,-1;\s2,-3;\s1,-2;\D3;\s-1,-2;\s-1,-1;\s-3,-1;\r4;\s-4,1;\s-1,1;\s-1,2;
\U3;\s1,2;\s2,3;\s8,2;\s2,1;\s1,2;\U2;\s-1,2;\s-3,1;\r4;\m132,858;\s3,1;\s3,4;\D23;\m170,170;\h681;\m170,193;\D23;\m238,182;\D12;
\m306,193;\D23;\m375,182;\D12;\m443,193;\D23;\m511,182;\D12;\m579,193;\D23;\m647,182;\D12;\m715,193;\D23;\m783,182;\D12;\m851,193;
\D23;\m169,148;\s-3,-2;\s-2,-3;\s-1,-5;\D3;\s1,-6;\s2,-3;\s3,-1;\h2;\s4,1;\s2,3;\s1,6;\U3;\s-1,5;\s-2,3;\s-4,2;\r2;\m302,143;\s2,1;
\s4,4;\D23;\m436,142;\U1;\s1,2;\s3,3;\h5;\s3,-3;\s1,-2;\D2;\s-1,-2;\s-2,-3;\s-11,-11;\h15;\m573,148;\h12;\s-6,-9;\h3;\s2,-1;\s1,-1;
\s1,-4;\D2;\s-1,-3;\s-2,-2;\s-3,-1;\r3;\s-4,1;\s-1,1;\s-1,2;\m718,148;\s-11,-16;\h17;\m718,148;\D23;\m857,148;\r11;\s-1,-10;\s1,1;
\s3,1;\h3;\s4,-1;\s2,-2;\s1,-4;\D2;\s-1,-3;\s-2,-2;\s-4,-1;\r3;\s-3,1;\s-1,1;\s-1,2;\m851,170;\U681;\m829,170;\h22;\m840,238;\h11;
\m829,306;\h22;\m840,375;\h11;\m829,443;\h22;\m840,511;\h11;\m829,579;\h22;\m840,647;\h11;\m829,715;\h22;\m840,783;\h11;\m829,851;
\h22;\m170,851;\h681;\m170,829;\U22;\m238,840;\U11;\m306,829;\U22;\m375,840;\U11;\m443,829;\U22;\m511,840;\U11;\m579,829;\U22;
\m647,840;\U11;\m715,829;\U22;\m783,840;\U11;\m851,829;\U22;\m184,340;\s3,42;\s4,41;\s3,37;\s3,34;\s5,43;\s3,31;\s3,26;\s3,28;
\s4,27;\s3,25;\s6,40;\s1,5;\s6,36;\s4,15;\s3,16;\s4,11;\s3,12;\s4,10;\s3,7;\s8,14;\s4,5;\s2,1;\s4,3;\s5,2;\h4;\s6,-1;\s6,-4;\h2;
\s7,-6;\s6,-7;\s10,-11;\s4,-6;\s14,-20;\s13,-21;\s14,-22;\s14,-21;\s13,-21;\s14,-20;\s13,-19;\s14,-19;\s14,-17;\s13,-17;\s28,-30;
\s13,-14;\s14,-13;\s13,-13;\s14,-11;\s14,-12;\s13,-10;\s14,-10;\s13,-10;\s14,-9;\s14,-8;\s13,-8;\s14,-8;\s14,-7;\s13,-7;\s14,-7;
\s13,-7;\s28,-12;\s13,-5;\s14,-6;\s14,-5;\s13,-5;\s14,-4;\s13,-5;\s14,-4;\s14,-5;\s13,-4;\m184,337;\s4,44;\s3,41;\s4,41;\s2,24;
\s5,43;\s3,31;\s3,28;\s3,22;\s4,28;\s3,25;\s4,21;\s6,38;\s4,18;\s4,16;\s2,11;\s8,24;\s4,10;\s2,4;\s4,7;\s5,6;\s5,5;\s5,3;\s6,1;\h2;
\s6,-2;\s8,-5;\s8,-6;\s5,-7;\s6,-8;\s8,-14;\s14,-26;\s13,-27;\s28,-56;\s13,-27;\s14,-27;\s13,-26;\s14,-25;\s14,-24;\s13,-23;
\s14,-23;\s14,-21;\s13,-20;\s14,-19;\s13,-18;\s14,-18;\s14,-16;\s13,-16;\s14,-15;\s13,-14;\s28,-26;\s13,-11;\s28,-22;\s13,-10;
\s14,-9;\s13,-8;\s28,-16;\s13,-7;\s28,-12;\s13,-5;\s14,-5;\s13,-5;\s28,-8;\s13,-3;\m184,324;\s3,36;\s4,36;\s3,31;\s3,29;\s4,32;
\s4,27;\s3,24;\s3,23;\s4,22;\s3,20;\s3,18;\s4,17;\s3,14;\s4,15;\s4,12;\s2,9;\s8,18;\s5,8;\s1,2;\s4,5;\s5,4;\s5,2;\s5,1;\s6,-1;
\s2,-1;\s7,-4;\s7,-5;\s8,-7;\s3,-3;\s2,-5;\s3,-6;\s11,-36;\s14,-44;\s13,-44;\s14,-44;\s14,-42;\s11,-32;\s2,-7;\s11,-29;\s3,-8;
\s10,-25;\s3,-9;\s10,-22;\s14,-28;\s4,-7;\s10,-19;\s3,-6;\s11,-16;\s3,-5;\s11,-15;\s3,-3;\s11,-13;\s2,-3;\s12,-12;\s2,-1;\s13,-12;
\s14,-9;\s14,-8;\s13,-6;\s14,-5;\s13,-4;\s28,-6;\s13,-1;\s14,-2;\s14,-1;\s13,-1;\h14;\s13,-1;\h14;\m184,302;\s6,48;\s3,22;\s3,21;
\s1,11;\s4,20;\s3,20;\s6,32;\s1,5;\s3,14;\s4,14;\s3,11;\s3,12;\s1,1;\s3,9;\s4,9;\s3,7;\s3,6;\s4,5;\s4,4;\s5,4;\s1,1;\s4,1;\s5,1;
\s10,-2;\s6,-4;\s2,-1;\s6,-5;\s8,-7;\s8,-6;\s2,-3;\s3,-6;\s2,-6;\s2,-8;\s4,-13;\s5,-25;\s1,-3;\s10,-37;\s10,-40;\s7,-24;\s6,-20;
\s7,-19;\s1,-4;\s6,-17;\s6,-16;\s2,-4;\s6,-14;\s6,-15;\s7,-14;\s8,-13;\s6,-11;\s7,-10;\s8,-9;\s6,-7;\s8,-8;\s6,-5;\s9,-7;\s4,-3;
\s14,-7;\s11,-4;\s3,-1;\s13,-4;\s14,-2;\s13,-2;\s14,-2;\h14;\s13,-1;\h14;
\fi\end{picture}
\end{center}
\caption{Radio pulse spectrum after vertical propagation from a given depth
(the maximum value is equal to 43.5 nV/MHz/TeV).}
\end{figure}
\vspace{0.5cm}

\begin{figure}[b]
\begin{center}
\begin{picture}(800,800)(0,0)\ifx\higzdraft\undefined\newcount\higzdraft\higzdraft=0{}\fi\ifnum\higzdraft>0
\put(0,0){\framebox(800,800){}}\else\ifx\higzstep\undefined\newcount\higzstep\higzstep=0{}\fi\ifnum\higzstep<1\higzstep=2
\fi\ifx\higzxx\undefined\newcount\higzxx\newcount\higzyy\newcount\higzx\newcount\higzy\newcount\higzdx\newcount\higzdy
\newcount\higzlx\newcount\higzly\newcount\higzslope\newcount\higzlen\newcount\higzllen\newcount\higzoffs\newcount\higzloffs
\newcount\higzadash\newcount\higzbdash\newcount\higzcdash\newcount\higzddash\newcount\higzmsize\newcount\higztemp\fi
\def\higzstroke#1,#2,#3,#4;{\advance\higzloffs\higzllen\ifnum\higzloffs>#1\advance\higzloffs-\higzllen\advance\higzloffs-#1
\higzloffs=-\higzloffs\ifnum#2>0\put(\higzlx,\higzly){\line(#3,#4){\higzloffs}}\fi\ifnum#2<0\put(\higzlx,\higzly){\circle*{0}}\fi
\higztemp=\higzloffs\multiply\higztemp#3\advance\higzlx\higztemp\higztemp=\higzloffs\multiply\higztemp#4\advance\higzly\higztemp
\advance\higzllen-\higzloffs\higzloffs=#1\else\ifnum#2>0\put(\higzlx,\higzly){\line(#3,#4){\higzllen}}\fi\ifnum#2<0\put(
\higzlx,\higzly){\circle*{0}}\fi\higzllen=0\fi}\def\higzdashed#1,#2,#3,#4,#5;{{\higzlx=#1\higzly=#2\higzllen=#5\higzloffs=
\higzoffs\loop\ifnum\higzloffs<\higzadash\ifnum\higzadash>1\higzstroke\higzadash,1,#3,#4;\else\higzstroke\higzadash,-1,#3,#4;\fi
\else\ifnum\higzloffs<\higzbdash\higzstroke\higzbdash,0,#3,#4;\else\ifnum\higzloffs<\higzcdash\higztemp=\higzcdash\advance
\higztemp-\higzbdash\ifnum\higztemp>1\higzstroke\higzcdash,1,#3,#4;\else\higzstroke\higzcdash,-1,#3,#4;\fi\else\ifnum\higzloffs<
\higzddash\higzstroke\higzddash,0,#3,#4;\else\higzloffs=0\fi\fi\fi\fi\ifnum\higzllen>0\repeat\global\higzoffs=\higzloffs}}\def
\higzsolid#1,#2,#3,#4,#5;{\put(#1,#2){\line(#3,#4){#5}}}\def\higzhslant#1,#2,#3;{\higzslope=#1\multiply\higzslope1000\advance
\higzslope500\divide\higzslope#2\higzlen=\higzslope\multiply\higzlen\higzstep\divide\higzlen1000\higzdy=0\loop\ifnum
\higzdy<#2\higzx=\higzxx\higzy=\higzyy\higzdx=\higzslope\multiply\higzdx\higzdy\advance\higzdx500\divide\higzdx1000\advance
\higzy\higzdy\multiply\higzdx#3\advance\higzx\higzdx\multiply\higzdx#3\advance\higzdx\higzlen\ifnum\higzdx>#1\advance
\higzlen#1\advance\higzlen-\higzdx\fi\higzline\higzx,\higzy,#3,0,\higzlen;\advance\higzdy\higzstep\repeat}\def\higzvslant#1,#2,#3;{
\higzslope=#2\multiply\higzslope1000\advance\higzslope500\divide\higzslope#1\higzlen=\higzslope\multiply\higzlen\higzstep
\divide\higzlen1000\higzdx=0\loop\ifnum\higzdx<#1\higzx=\higzxx\higzy=\higzyy\higzdy=\higzslope\multiply\higzdy\higzdx\advance
\higzdy500\divide\higzdy1000\advance\higzx\higzdx\multiply\higzdy#3\advance\higzy\higzdy\multiply\higzdy#3\advance
\higzdy\higzlen\ifnum\higzdy>#2\advance\higzlen#2\advance\higzlen-\higzdy\fi\higzline\higzx,\higzy,0,#3,\higzlen;\advance\higzdx
\higzstep\repeat}\def\s#1,#2;{\higzdx=#1{}\ifnum\higzdx<0\higzdx=-\higzdx\fi\higzdy=#2{}\ifnum\higzdy<0\higzdy=-\higzdy\fi\ifnum
\higzdx<\higzdy\ifnum#1<0\advance\higzxx#1\advance\higzyy#2\ifnum#2<0\higzvslant-#1,-#2,1;\else\higzvslant-#1,#2,-1;
\fi\else\ifnum#2<0\higzvslant#1,-#2,-1;\else\higzvslant#1,#2,1;\fi\advance\higzxx#1\advance\higzyy#2\fi\else\ifnum#2<0
\advance\higzxx#1\advance\higzyy#2\ifnum#1<0\higzhslant-#1,-#2,1;\else\higzhslant#1,-#2,-1;\fi\else\ifnum#1<0
\higzhslant-#1,#2,-1;\else\higzhslant#1,#2,1;\fi\advance\higzxx#1\advance\higzyy#2\fi\fi}\def\h#1;{\higzline\higzxx,
\higzyy,1,0,#1;\advance\higzxx#1}\def\r#1;{\higzline\higzxx,\higzyy,-1,0,#1;\advance\higzxx-#1}\def\U#1;{\higzline\higzxx,
\higzyy,0,1,#1;\advance\higzyy#1}\def\D#1;{\higzline\higzxx,\higzyy,0,-1,#1;\advance\higzyy-#1}\def\m#1,#2;{\higzxx=#1
\higzyy=#2}\def\higzdot#1,#2;{\put(#1,#2){\circle*{\higzmsize}}}\def\higzplus#1,#2;{\higzx=#1\multiply\higzx2\advance\higzx-
\higzmsize\divide\higzx2\put(\higzx,#2){\line(1,0){\higzmsize}}\higzy=#2\multiply\higzy2\advance\higzy-\higzmsize\divide
\higzy2\put(#1,\higzy){\line(0,1){\higzmsize}}}\def\higzstar#1,#2;{\higzplus#1,#2;\higzcross#1,#2;}\def
\higzcircle#1,#2;{\put(#1,#2){\circle{\higzmsize}}}\def\higzcross#1,#2;{\let\higzsave\higzline\let\higzline\higzsolid\higzlx=#1
\multiply\higzlx2\advance\higzlx-\higzmsize\divide\higzlx2\higzly=#2\multiply\higzly2\advance\higzly-\higzmsize\divide
\higzly2\m\higzlx,\higzly;\s\higzmsize,\higzmsize;\higzly=#2\multiply\higzly2\advance\higzly\higzmsize\divide\higzly2\m\higzlx,
\higzly;\s\higzmsize,-\higzmsize;\let\higzline\higzsave}\def\p#1,#2;{\higzmarker#1,#2;}\def\f#1,#2;{\put(\higzxx,
\higzyy){\makebox(#1,#2)[lb]{\rule{#1\unitlength}{#2\unitlength}}}}
\let\higzline=\higzsolid\let\higzline=\higzsolid\m686,89;\D23;\m686,89;\h7;\s4,-2;\s2,-2;\s1,-2;\s1,-3;\D6;\s-1,-3;\s-1,-2;\s-2,-2;
\s-4,-1;\r7;\m707,74;\h13;\U3;\s-1,2;\s-1,1;\s-2,1;\r3;\s-2,-1;\s-2,-2;\s-2,-4;\D2;\s2,-3;\s2,-2;\s2,-1;\h3;\s2,1;\s2,2;\m728,81;
\D23;\m728,78;\s2,2;\s2,1;\h4;\s2,-1;\s2,-2;\s1,-4;\D2;\s-1,-3;\s-2,-2;\s-2,-1;\r4;\s-2,1;\s-2,2;\m750,89;\D19;\s1,-3;\s2,-1;\h2;
\m746,81;\h8;\m762,89;\D23;\m762,77;\s3,3;\s2,1;\h3;\s2,-1;\s1,-3;\D11;\m800,93;\s-2,-2;\s-4,-8;\s-1,-5;\D5;\s1,-5;\s4,-8;\s2,-2;
\m808,81;\D15;\m808,77;\s3,3;\s2,1;\h4;\s2,-1;\s1,-3;\D11;\m820,77;\s3,3;\s2,1;\h4;\s2,-1;\s1,-3;\D11;\m839,93;\s3,-2;\s4,-8;\s1,-5;
\D5;\s-1,-5;\s-4,-8;\s-3,-2;\m34,437;\h23;\m34,437;\U10;\s1,3;\s1,1;\s2,1;\h3;\s2,-1;\s1,-1;\s1,-3;\D10;\m45,445;\s12,7;\m48,459;
\U13;\r2;\s-2,-1;\s-1,-1;\s-1,-2;\D4;\s1,-2;\s2,-2;\s3,-1;\h2;\s4,1;\s2,2;\s1,2;\U4;\s-1,2;\s-2,2;\m34,479;\h23;\m42,500;\h15;
\m45,500;\s-2,-2;\s-1,-2;\D4;\s1,-2;\s2,-2;\s3,-1;\h2;\s4,1;\s2,2;\s1,2;\U4;\s-1,2;\s-2,2;\m34,510;\h18;\s4,1;\s1,2;\U2;\m42,506;
\U8;\m34,520;\s1,2;\s-1,1;\s-1,-1;\s1,-2;\m42,522;\h15;\m42,528;\s15,7;\m42,541;\s15,-6;\m48,546;\U13;\r2;\s-2,-1;\s-1,-1;\s-1,-2;
\D3;\s1,-2;\s2,-2;\s3,-2;\h2;\s4,2;\s2,2;\s1,2;\U3;\s-1,2;\s-2,2;\m34,579;\h18;\s4,1;\s1,2;\U2;\m42,576;\U7;\m34,591;\h23;\m46,591;
\s-3,3;\s-1,2;\U3;\s1,3;\s3,1;\h11;\m42,611;\h15;\m48,611;\s-3,1;\s-2,3;\s-1,2;\U3;\m48,624;\U13;\r2;\s-2,-1;\s-1,-1;\s-1,-2;\D3;
\s1,-3;\s2,-2;\s3,-1;\h2;\s4,1;\s2,2;\s1,3;\U3;\s-1,2;\s-2,2;\m45,656;\s-2,-1;\s-1,-4;\D3;\s1,-3;\s2,-1;\s2,1;\s1,2;\s1,5;\s1,3;
\s2,1;\h2;\s2,-1;\s1,-4;\D3;\s-1,-3;\s-2,-1;\m34,663;\h23;\m46,663;\s-3,3;\s-1,3;\U3;\s1,2;\s3,1;\h11;\m42,688;\s1,-2;\s2,-2;\s3,-1;
\h2;\s4,1;\s2,2;\s1,2;\U3;\s-1,2;\s-2,3;\s-4,1;\r2;\s-3,-1;\s-2,-3;\s-1,-2;\D3;\m34,704;\h23;\m34,725;\h23;\m45,725;\s-2,-2;\s-1,-3;
\D3;\s1,-2;\s2,-2;\s3,-1;\h2;\s4,1;\s2,2;\s1,2;\U3;\s-1,3;\s-2,2;\m48,743;\U13;\r2;\s-2,-1;\s-1,-1;\s-1,-2;\D3;\s1,-3;\s2,-2;\s3,-1;
\h2;\s4,1;\s2,2;\s1,3;\U3;\s-1,2;\s-2,2;\m42,764;\h15;\m46,764;\s-3,3;\s-1,2;\U3;\s1,3;\s3,1;\h11;\m48,783;\U13;\r2;\s-2,-1;\s-1,-1;
\s-1,-2;\D3;\s1,-3;\s2,-2;\s3,-1;\h2;\s4,1;\s2,2;\s1,3;\U3;\s-1,2;\s-2,2;\m42,804;\h15;\m48,804;\s-3,1;\s-2,2;\s-1,2;\U3;\m42,830;
\h17;\s3,-1;\s2,-4;\D3;\s-1,-2;\m45,830;\s-2,-3;\s-1,-2;\D3;\s1,-2;\s2,-2;\s3,-1;\h2;\s4,1;\s2,2;\s1,2;\U3;\s-1,2;\s-2,3;\m42,836;
\s15,7;\m42,849;\s15,-6;\s4,-3;\s2,-2;\s1,-2;\D1;\m507,294;\s5,-15;\m516,294;\s-4,-15;\m516,294;\s4,-15;\m524,294;\s-4,-15;
\m531,302;\s1,-1;\s1,1;\s-1,1;\s-1,-1;\m532,294;\D15;\m542,302;\D18;\s1,-4;\s2,-1;\h2;\m539,294;\h7;\m554,302;\D23;\m554,290;\s3,3;
\s2,1;\h3;\s2,-1;\s2,-3;\D11;\m579,294;\s-3,-1;\s-2,-2;\s-1,-3;\D2;\s1,-4;\s2,-2;\s3,-1;\h3;\s2,1;\s2,2;\s1,4;\U2;\s-1,3;\s-2,2;
\s-2,1;\r3;\m595,294;\D10;\s1,-4;\s2,-1;\h3;\s2,1;\s4,4;\m607,294;\D15;\m616,302;\D18;\s1,-4;\s3,-1;\h2;\m613,294;\h8;\m651,294;
\D15;\m651,291;\s-2,2;\s-2,1;\r4;\s-2,-1;\s-2,-2;\s-1,-3;\D2;\s1,-4;\s2,-2;\s2,-1;\h4;\s2,1;\s2,2;\m660,302;\D23;\m660,291;\s2,2;
\s2,1;\h3;\s2,-1;\s2,-2;\s2,-3;\D2;\s-2,-4;\s-2,-2;\s-2,-1;\r3;\s-2,1;\s-2,2;\m691,291;\s-1,2;\s-3,1;\r4;\s-3,-1;\s-1,-2;\s1,-2;
\s2,-1;\s6,-1;\s2,-1;\s1,-2;\D2;\s-1,-2;\s-3,-1;\r4;\s-3,1;\s-1,2;\m703,294;\s-2,-1;\s-2,-2;\s-2,-3;\D2;\s2,-4;\s2,-2;\s2,-1;\h3;
\s2,1;\s2,2;\s1,4;\U2;\s-1,3;\s-2,2;\s-2,1;\r3;\m719,294;\D15;\m719,288;\s1,3;\s2,2;\s2,1;\h4;\m733,294;\D22;\m733,291;\s2,2;\s2,1;
\h4;\s2,-1;\s2,-2;\s1,-3;\D2;\s-1,-4;\s-2,-2;\s-2,-1;\r4;\s-2,1;\s-2,2;\m755,302;\D18;\s1,-4;\s2,-1;\h2;\m751,294;\h8;\m766,302;
\s1,-1;\s1,1;\s-1,1;\s-1,-1;\m767,294;\D15;\m780,294;\s-3,-1;\s-2,-2;\s-1,-3;\D2;\s1,-4;\s2,-2;\s3,-1;\h3;\s2,1;\s2,2;\s1,4;\U2;
\s-1,3;\s-2,2;\s-2,1;\r3;\m796,294;\D15;\m796,290;\s3,3;\s2,1;\h3;\s3,-1;\s1,-3;\D11;\m445,621;\D15;\m445,615;\s1,3;\s2,2;\s3,1;\h3;
\m458,615;\h13;\U2;\s-1,2;\s-1,1;\s-2,1;\r3;\s-3,-1;\s-2,-2;\s-1,-3;\D2;\s1,-4;\s2,-2;\s3,-1;\h3;\s2,1;\s2,2;\m491,621;\D15;
\m491,618;\s-3,2;\s-2,1;\r3;\s-2,-1;\s-2,-2;\s-1,-3;\D2;\s1,-4;\s2,-2;\s2,-1;\h3;\s2,1;\s3,2;\m499,629;\D23;\m531,621;\D15;
\m531,618;\s-3,2;\s-2,1;\r3;\s-2,-1;\s-2,-2;\s-1,-3;\D2;\s1,-4;\s2,-2;\s2,-1;\h3;\s2,1;\s3,2;\m539,621;\D15;\m539,617;\s3,3;\s3,1;
\h3;\s2,-1;\s1,-3;\D11;\m561,629;\D19;\s1,-3;\s2,-1;\h2;\m558,621;\h7;\m585,621;\D15;\m585,618;\s-3,2;\s-2,1;\r3;\s-2,-1;\s-2,-2;
\s-1,-3;\D2;\s1,-4;\s2,-2;\s2,-1;\h3;\s2,1;\s3,2;\m593,621;\D15;\m593,615;\s1,3;\s2,2;\s3,1;\h3;\m619,618;\s-2,2;\s-2,1;\r3;\s-3,-1;
\s-2,-2;\s-1,-3;\D2;\s1,-4;\s2,-2;\s3,-1;\h3;\s2,1;\s2,2;\m628,629;\D19;\s1,-3;\s2,-1;\h2;\m625,621;\h7;\m639,629;\s1,-1;\s1,1;
\s-1,1;\s-1,-1;\m640,621;\D15;\m660,618;\s-2,2;\s-2,1;\r3;\s-2,-1;\s-3,-2;\s-1,-3;\D2;\s1,-4;\s3,-2;\s2,-1;\h3;\s2,1;\s2,2;
\m678,629;\s1,-1;\s1,1;\s-1,1;\s-1,-1;\m679,621;\D15;\m699,618;\s-2,2;\s-2,1;\r3;\s-3,-1;\s-2,-2;\s-1,-3;\D2;\s1,-4;\s2,-2;\s3,-1;
\h3;\s2,1;\s2,2;\m706,615;\h13;\U2;\s-1,2;\s-4,2;\r3;\s-2,-1;\s-2,-2;\s-1,-3;\D2;\s1,-4;\s2,-2;\s2,-1;\h3;\s2,1;\s3,2;\put(170,170)
{\framebox(681,681){}}\m170,170;\U681;\m193,170;\r23;\m182,197;\r12;\m182,225;\r12;\m182,252;\r12;\m182,279;\r12;\m193,306;\r23;
\m182,334;\r12;\m182,361;\r12;\m182,388;\r12;\m182,415;\r12;\m193,443;\r23;\m182,470;\r12;\m182,497;\r12;\m182,524;\r12;\m182,552;
\r12;\m193,579;\r23;\m182,606;\r12;\m182,633;\r12;\m182,661;\r12;\m182,688;\r12;\m193,715;\r23;\m182,742;\r12;\m182,770;\r12;
\m182,797;\r12;\m182,824;\r12;\m193,851;\r23;\m136,182;\s-4,-1;\s-2,-4;\s-1,-5;\D3;\s1,-6;\s2,-3;\s4,-1;\h2;\s3,1;\s2,3;\s1,6;\U3;
\s-1,5;\s-2,4;\s-3,1;\r2;\m142,318;\r11;\s-1,-10;\s1,1;\s4,1;\h3;\s3,-1;\s2,-2;\s1,-3;\D2;\s-1,-4;\s-2,-2;\s-3,-1;\r3;\s-4,1;\s-1,1;
\s-1,2;\m111,450;\s2,1;\s3,3;\D23;\m136,454;\s-4,-1;\s-2,-3;\s-1,-6;\D3;\s1,-5;\s2,-4;\s4,-1;\h2;\s3,1;\s2,4;\s1,5;\U3;\s-1,6;
\s-2,3;\s-3,1;\r2;\m111,586;\s2,1;\s3,3;\D22;\m142,590;\r11;\s-1,-10;\s1,2;\s4,1;\h3;\s3,-1;\s2,-3;\s1,-3;\D2;\s-1,-3;\s-2,-2;
\s-3,-1;\r3;\s-4,1;\s-1,1;\s-1,2;\m109,721;\U1;\s1,2;\s1,1;\s2,1;\h4;\s4,-2;\s1,-2;\D2;\s-1,-2;\s-3,-3;\s-10,-11;\h15;\m136,726;
\s-4,-1;\s-2,-3;\s-1,-5;\D4;\s1,-5;\s2,-3;\s4,-1;\h2;\s3,1;\s2,3;\s1,5;\U4;\s-1,5;\s-2,3;\s-3,1;\r2;\m109,857;\U1;\s2,4;\s2,1;\h4;
\s2,-1;\s2,-2;\s1,-2;\D2;\s-1,-2;\s-3,-3;\s-10,-11;\h15;\m142,863;\r11;\s-1,-10;\s1,1;\s4,1;\h3;\s3,-1;\s2,-2;\s1,-3;\D3;\s-1,-3;
\s-2,-2;\s-3,-1;\r3;\s-4,1;\s-1,1;\s-1,2;\m170,170;\h681;\m170,193;\D23;\m238,182;\D12;\m306,193;\D23;\m375,182;\D12;\m443,193;\D23;
\m511,182;\D12;\m579,193;\D23;\m647,182;\D12;\m715,193;\D23;\m783,182;\D12;\m851,193;\D23;\m169,148;\s-3,-2;\s-2,-3;\s-1,-5;\D3;
\s1,-6;\s2,-3;\s3,-1;\h2;\s4,1;\s2,3;\s1,6;\U3;\s-1,5;\s-2,3;\s-4,2;\r2;\m278,142;\U1;\s1,2;\s2,1;\s2,2;\h4;\s3,-3;\s1,-2;\D2;
\s-1,-2;\s-2,-3;\s-11,-11;\h15;\m305,148;\s-3,-2;\s-2,-3;\s-1,-5;\D3;\s1,-6;\s2,-3;\s3,-1;\h3;\s3,1;\s2,3;\s1,6;\U3;\s-1,5;\s-2,3;
\s-3,2;\r3;\m327,148;\s-3,-2;\s-2,-3;\s-1,-5;\D3;\s1,-6;\s2,-3;\s3,-1;\h2;\s3,1;\s3,3;\s1,6;\U3;\s-1,5;\s-3,3;\s-3,2;\r2;\m424,148;
\s-11,-16;\h17;\m424,148;\D23;\m442,148;\s-4,-2;\s-2,-3;\s-1,-5;\D3;\s1,-6;\s2,-3;\s4,-1;\h2;\s3,1;\s2,3;\s1,6;\U3;\s-1,5;\s-2,3;
\s-3,2;\r2;\m463,148;\s-3,-2;\s-2,-3;\s-1,-5;\D3;\s1,-6;\s2,-3;\s3,-1;\h2;\s4,1;\s2,3;\s1,6;\U3;\s-1,5;\s-2,3;\s-4,2;\r2;\m564,144;
\s-1,2;\s-4,2;\r2;\s-3,-2;\s-2,-3;\s-1,-5;\D6;\s1,-4;\s2,-2;\s3,-1;\h1;\s4,1;\s2,2;\s1,3;\U1;\s-1,4;\s-2,2;\s-4,1;\r1;\s-3,-1;
\s-2,-2;\s-1,-4;\m578,148;\s-3,-2;\s-3,-3;\s-1,-5;\D3;\s1,-6;\s3,-3;\s3,-1;\h2;\s3,1;\s2,3;\s1,6;\U3;\s-1,5;\s-2,3;\s-3,2;\r2;
\m599,148;\s-3,-2;\s-2,-3;\s-1,-5;\D3;\s1,-6;\s2,-3;\s3,-1;\h3;\s3,1;\s2,3;\s1,6;\U3;\s-1,5;\s-2,3;\s-3,2;\r3;\m691,148;\s-3,-2;
\s-1,-2;\D2;\s1,-2;\s2,-1;\s5,-1;\s3,-1;\s2,-2;\s1,-3;\D3;\s-1,-2;\s-1,-1;\s-3,-1;\r5;\s-3,1;\s-1,1;\s-1,2;\U3;\s1,3;\s2,2;\s3,1;
\s5,1;\s2,1;\s1,2;\U2;\s-1,2;\s-3,2;\r5;\m714,148;\s-3,-2;\s-2,-3;\s-2,-5;\D3;\s2,-6;\s2,-3;\s3,-1;\h2;\s3,1;\s3,3;\s1,6;\U3;\s-1,5;
\s-3,3;\s-3,2;\r2;\m736,148;\s-4,-2;\s-2,-3;\s-1,-5;\D3;\s1,-6;\s2,-3;\s4,-1;\h2;\s3,1;\s2,3;\s1,6;\U3;\s-1,5;\s-2,3;\s-3,2;\r2;
\m814,143;\s3,1;\s3,4;\D23;\m839,148;\s-3,-2;\s-2,-3;\s-1,-5;\D3;\s1,-6;\s2,-3;\s3,-1;\h3;\s3,1;\s2,3;\s1,6;\U3;\s-1,5;\s-2,3;
\s-3,2;\r3;\m861,148;\s-3,-2;\s-2,-3;\s-2,-5;\D3;\s2,-6;\s2,-3;\s3,-1;\h2;\s3,1;\s3,3;\s1,6;\U3;\s-1,5;\s-3,3;\s-3,2;\r2;\m883,148;
\s-4,-2;\s-2,-3;\s-1,-5;\D3;\s1,-6;\s2,-3;\s4,-1;\h2;\s3,1;\s2,3;\s1,6;\U3;\s-1,5;\s-2,3;\s-3,2;\r2;\m851,170;\U681;\m829,170;\h22;
\m840,197;\h11;\m840,225;\h11;\m840,252;\h11;\m840,279;\h11;\m829,306;\h22;\m840,334;\h11;\m840,361;\h11;\m840,388;\h11;\m840,415;
\h11;\m829,443;\h22;\m840,470;\h11;\m840,497;\h11;\m840,524;\h11;\m840,552;\h11;\m829,579;\h22;\m840,606;\h11;\m840,633;\h11;
\m840,661;\h11;\m840,688;\h11;\m829,715;\h22;\m840,742;\h11;\m840,770;\h11;\m840,797;\h11;\m840,824;\h11;\m829,851;\h22;\m170,851;
\h681;\m170,829;\U22;\m238,840;\U11;\m306,829;\U22;\m375,840;\U11;\m443,829;\U22;\m511,840;\U11;\m579,829;\U22;\m647,840;\U11;
\m715,829;\U22;\m783,840;\U11;\m851,829;\U22;\m177,173;\s7,3;\s7,2;\s6,3;\s14,6;\s7,2;\s14,6;\s6,2;\s26,12;\s42,22;\s37,20;\s32,20;
\s34,22;\s34,24;\s32,24;\s36,30;\s29,26;\s29,27;\s10,9;\s27,28;\s26,28;\s15,17;\s25,30;\s24,29;\s19,26;\s24,33;\s20,30;\s21,32;
\s3,5;\s23,39;\s21,35;\s24,46;\m238,197;\s9,4;\m255,204;\s9,4;\m272,211;\s9,3;\m289,218;\s8,3;\m306,224;\U1;\m306,225;\s8,3;
\m323,231;\s8,4;\m340,238;\s8,3;\m356,245;\s9,3;\m373,251;\s2,1;\m375,252;\s7,3;\m390,258;\s9,4;\m407,265;\s8,3;\m424,272;\s8,3;
\m441,278;\s2,1;\m443,279;\s6,3;\m458,285;\s8,4;\m474,292;\s9,3;\m491,299;\s9,3;\m508,305;\s3,1;\m511,306;\s6,3;\m525,312;\s8,4;
\m542,319;\s8,3;\m559,326;\s8,3;\m576,332;\s3,2;\m579,334;\s5,2;\m592,339;\s9,3;\m609,346;\s9,3;\m626,353;\s9,3;\m643,359;\s4,2;
\m647,361;\s4,2;\m660,366;\s8,3;\m677,373;\s8,3;\m694,380;\s8,3;\m710,386;\s5,2;\m715,388;\s4,2;\m727,393;\s9,3;\m744,400;\s9,3;
\m761,407;\s8,3;\m778,413;\s5,2;\m783,415;\s3,2;\m795,420;\s8,3;\m812,427;\s8,3;\m828,434;\s9,3;\m845,440;\s6,3;
\fi\end{picture}
\end{center}
\caption{Relative threshold dependence on depth ($E_{100}$ is taken to be
unity).}
\end{figure}
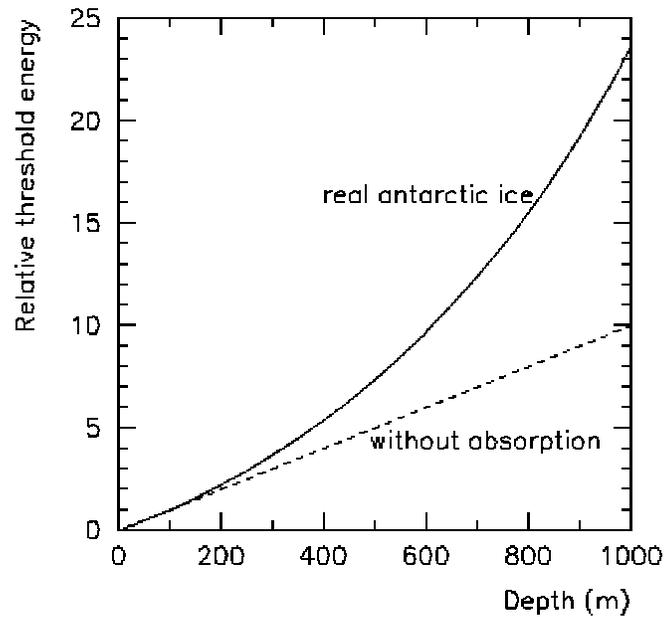
\clearpage

\begin{figure}[t]
\begin{center}
\begin{picture}(800,800)(0,0)\ifx\higzdraft\undefined\newcount\higzdraft\higzdraft=0{}\fi\ifnum\higzdraft>0
\put(0,0){\framebox(800,800){}}\else\ifx\higzstep\undefined\newcount\higzstep\higzstep=0{}\fi\ifnum\higzstep<1\higzstep=2
\fi\ifx\higzxx\undefined\newcount\higzxx\newcount\higzyy\newcount\higzx\newcount\higzy\newcount\higzdx\newcount\higzdy
\newcount\higzlx\newcount\higzly\newcount\higzslope\newcount\higzlen\newcount\higzllen\newcount\higzoffs\newcount\higzloffs
\newcount\higzadash\newcount\higzbdash\newcount\higzcdash\newcount\higzddash\newcount\higzmsize\newcount\higztemp\fi
\def\higzstroke#1,#2,#3,#4;{\advance\higzloffs\higzllen\ifnum\higzloffs>#1\advance\higzloffs-\higzllen\advance\higzloffs-#1
\higzloffs=-\higzloffs\ifnum#2>0\put(\higzlx,\higzly){\line(#3,#4){\higzloffs}}\fi\ifnum#2<0\put(\higzlx,\higzly){\circle*{0}}\fi
\higztemp=\higzloffs\multiply\higztemp#3\advance\higzlx\higztemp\higztemp=\higzloffs\multiply\higztemp#4\advance\higzly\higztemp
\advance\higzllen-\higzloffs\higzloffs=#1\else\ifnum#2>0\put(\higzlx,\higzly){\line(#3,#4){\higzllen}}\fi\ifnum#2<0\put(
\higzlx,\higzly){\circle*{0}}\fi\higzllen=0\fi}\def\higzdashed#1,#2,#3,#4,#5;{{\higzlx=#1\higzly=#2\higzllen=#5\higzloffs=
\higzoffs\loop\ifnum\higzloffs<\higzadash\ifnum\higzadash>1\higzstroke\higzadash,1,#3,#4;\else\higzstroke\higzadash,-1,#3,#4;\fi
\else\ifnum\higzloffs<\higzbdash\higzstroke\higzbdash,0,#3,#4;\else\ifnum\higzloffs<\higzcdash\higztemp=\higzcdash\advance
\higztemp-\higzbdash\ifnum\higztemp>1\higzstroke\higzcdash,1,#3,#4;\else\higzstroke\higzcdash,-1,#3,#4;\fi\else\ifnum\higzloffs<
\higzddash\higzstroke\higzddash,0,#3,#4;\else\higzloffs=0\fi\fi\fi\fi\ifnum\higzllen>0\repeat\global\higzoffs=\higzloffs}}\def
\higzsolid#1,#2,#3,#4,#5;{\put(#1,#2){\line(#3,#4){#5}}}\def\higzhslant#1,#2,#3;{\higzslope=#1\multiply\higzslope1000\advance
\higzslope500\divide\higzslope#2\higzlen=\higzslope\multiply\higzlen\higzstep\divide\higzlen1000\higzdy=0\loop\ifnum
\higzdy<#2\higzx=\higzxx\higzy=\higzyy\higzdx=\higzslope\multiply\higzdx\higzdy\advance\higzdx500\divide\higzdx1000\advance
\higzy\higzdy\multiply\higzdx#3\advance\higzx\higzdx\multiply\higzdx#3\advance\higzdx\higzlen\ifnum\higzdx>#1\advance
\higzlen#1\advance\higzlen-\higzdx\fi\higzline\higzx,\higzy,#3,0,\higzlen;\advance\higzdy\higzstep\repeat}\def\higzvslant#1,#2,#3;{
\higzslope=#2\multiply\higzslope1000\advance\higzslope500\divide\higzslope#1\higzlen=\higzslope\multiply\higzlen\higzstep
\divide\higzlen1000\higzdx=0\loop\ifnum\higzdx<#1\higzx=\higzxx\higzy=\higzyy\higzdy=\higzslope\multiply\higzdy\higzdx\advance
\higzdy500\divide\higzdy1000\advance\higzx\higzdx\multiply\higzdy#3\advance\higzy\higzdy\multiply\higzdy#3\advance
\higzdy\higzlen\ifnum\higzdy>#2\advance\higzlen#2\advance\higzlen-\higzdy\fi\higzline\higzx,\higzy,0,#3,\higzlen;\advance\higzdx
\higzstep\repeat}\def\s#1,#2;{\higzdx=#1{}\ifnum\higzdx<0\higzdx=-\higzdx\fi\higzdy=#2{}\ifnum\higzdy<0\higzdy=-\higzdy\fi\ifnum
\higzdx<\higzdy\ifnum#1<0\advance\higzxx#1\advance\higzyy#2\ifnum#2<0\higzvslant-#1,-#2,1;\else\higzvslant-#1,#2,-1;
\fi\else\ifnum#2<0\higzvslant#1,-#2,-1;\else\higzvslant#1,#2,1;\fi\advance\higzxx#1\advance\higzyy#2\fi\else\ifnum#2<0
\advance\higzxx#1\advance\higzyy#2\ifnum#1<0\higzhslant-#1,-#2,1;\else\higzhslant#1,-#2,-1;\fi\else\ifnum#1<0
\higzhslant-#1,#2,-1;\else\higzhslant#1,#2,1;\fi\advance\higzxx#1\advance\higzyy#2\fi\fi}\def\h#1;{\higzline\higzxx,
\higzyy,1,0,#1;\advance\higzxx#1}\def\r#1;{\higzline\higzxx,\higzyy,-1,0,#1;\advance\higzxx-#1}\def\U#1;{\higzline\higzxx,
\higzyy,0,1,#1;\advance\higzyy#1}\def\D#1;{\higzline\higzxx,\higzyy,0,-1,#1;\advance\higzyy-#1}\def\m#1,#2;{\higzxx=#1
\higzyy=#2}\def\higzdot#1,#2;{\put(#1,#2){\circle*{\higzmsize}}}\def\higzplus#1,#2;{\higzx=#1\multiply\higzx2\advance\higzx-
\higzmsize\divide\higzx2\put(\higzx,#2){\line(1,0){\higzmsize}}\higzy=#2\multiply\higzy2\advance\higzy-\higzmsize\divide
\higzy2\put(#1,\higzy){\line(0,1){\higzmsize}}}\def\higzstar#1,#2;{\higzplus#1,#2;\higzcross#1,#2;}\def
\higzcircle#1,#2;{\put(#1,#2){\circle{\higzmsize}}}\def\higzcross#1,#2;{\let\higzsave\higzline\let\higzline\higzsolid\higzlx=#1
\multiply\higzlx2\advance\higzlx-\higzmsize\divide\higzlx2\higzly=#2\multiply\higzly2\advance\higzly-\higzmsize\divide
\higzly2\m\higzlx,\higzly;\s\higzmsize,\higzmsize;\higzly=#2\multiply\higzly2\advance\higzly\higzmsize\divide\higzly2\m\higzlx,
\higzly;\s\higzmsize,-\higzmsize;\let\higzline\higzsave}\def\p#1,#2;{\higzmarker#1,#2;}\def\f#1,#2;{\put(\higzxx,
\higzyy){\makebox(#1,#2)[lb]{\rule{#1\unitlength}{#2\unitlength}}}}
\let\higzline=\higzsolid\let\higzline=\higzsolid\m655,89;\D23;\m655,66;\h13;\m678,81;\s-2,-1;\s-3,-2;\s-1,-4;\D2;\s1,-3;\s3,-2;
\s2,-1;\h3;\s2,1;\s2,2;\s1,3;\U2;\s-1,4;\s-2,2;\s-2,1;\r3;\m706,81;\D17;\s-1,-4;\s-1,-1;\s-2,-1;\r4;\s-2,1;\m706,78;\s-2,2;\s-2,1;
\r4;\s-2,-1;\s-2,-2;\s-1,-4;\D2;\s1,-3;\s2,-2;\s2,-1;\h4;\s2,1;\s2,2;\m722,93;\s-2,-2;\s-4,-8;\s-1,-5;\D5;\s1,-5;\s4,-8;\s2,-2;
\m730,89;\D23;\m730,89;\h14;\m730,78;\h8;\m730,66;\h14;\m750,68;\s-2,-2;\s-1,-2;\D1;\s3,-3;\h2;\s3,3;\U1;\s-1,2;\s-2,2;\r2;\m778,93;
\s-20,-35;\m789,89;\D23;\m781,89;\h15;\m800,74;\h13;\U3;\s-1,2;\s-1,1;\s-2,1;\r3;\s-2,-1;\s-2,-2;\s-2,-4;\D2;\s2,-3;\s2,-2;\s2,-1;
\h3;\s2,1;\s2,2;\m818,89;\s8,-23;\m835,89;\s-9,-23;\m839,93;\s3,-2;\s4,-8;\s1,-5;\D5;\s-1,-5;\s-4,-8;\s-3,-2;\m37,413;\s-2,-2;
\s-1,-3;\D4;\s1,-4;\s2,-2;\h2;\s3,1;\s1,1;\s3,9;\s1,2;\s1,1;\s2,1;\h4;\s2,-2;\s1,-3;\D4;\s-1,-4;\s-2,-2;\m34,421;\h23;\m46,421;
\s-3,3;\s-1,2;\U4;\s1,2;\s3,1;\h11;\m42,446;\s1,-2;\s2,-2;\s3,-2;\h2;\s4,2;\s2,2;\s1,2;\U3;\s-1,2;\s-2,2;\s-4,2;\r2;\s-3,-2;\s-2,-2;
\s-1,-2;\D3;\m42,461;\s15,4;\m42,470;\s15,-5;\m42,470;\s15,4;\m42,478;\s15,-4;\m48,485;\U13;\r2;\s-2,-1;\s-1,-1;\s-1,-3;\D3;\s1,-2;
\s2,-2;\s3,-1;\h2;\s4,1;\s2,2;\s1,2;\U3;\s-1,3;\s-2,2;\m42,505;\h15;\m48,505;\s-3,1;\s-2,3;\s-1,2;\U3;\m42,530;\h22;\m45,530;\s-2,2;
\s-1,3;\U3;\s1,2;\s2,2;\s3,1;\h2;\s4,-1;\s2,-2;\s1,-2;\D3;\s-1,-3;\s-2,-2;\m42,551;\h15;\m48,551;\s-3,1;\s-2,2;\s-1,2;\U3;\m42,569;
\s1,-2;\s2,-2;\s3,-1;\h2;\s4,1;\s2,2;\s1,2;\U3;\s-1,3;\s-2,2;\s-4,1;\r2;\s-3,-1;\s-2,-2;\s-1,-3;\D3;\m34,597;\h23;\m45,597;\s-2,-2;
\s-1,-2;\D3;\s1,-3;\s2,-2;\s3,-1;\h2;\s4,1;\s2,2;\s1,3;\U3;\s-1,2;\s-2,2;\m42,606;\h10;\s4,1;\s1,2;\U3;\s-1,3;\s-4,3;\m42,618;\h15;
\m45,638;\s-2,-2;\s-1,-2;\D3;\s1,-2;\s2,-3;\s3,-1;\h2;\s4,1;\s2,3;\s1,2;\U3;\s-1,2;\s-2,2;\m34,647;\h18;\s4,1;\s1,2;\U2;\m42,644;
\U7;\m34,658;\s1,1;\s-1,1;\s-1,-1;\s1,-1;\m42,659;\h15;\m42,672;\s1,-2;\s2,-3;\s3,-1;\h2;\s4,1;\s2,3;\s1,2;\U3;\s-1,2;\s-2,2;\s-4,1;
\r2;\s-3,-1;\s-2,-2;\s-1,-2;\D3;\m42,688;\h15;\m46,688;\s-3,3;\s-1,2;\U4;\s1,2;\s3,1;\h11;\m34,726;\D2;\s1,-2;\s3,-2;\h19;\m42,717;
\U8;\m42,732;\h10;\s4,1;\s1,3;\U3;\s-1,2;\s-4,3;\m42,744;\h15;\m42,753;\h15;\m46,753;\s-3,3;\s-1,2;\U4;\s1,2;\s3,1;\h11;\m45,785;
\s-2,-2;\s-1,-2;\D3;\s1,-2;\s2,-3;\s3,-1;\h2;\s4,1;\s2,3;\s1,2;\U3;\s-1,2;\s-2,2;\m34,794;\h18;\s4,1;\s1,2;\U2;\m42,791;\U7;
\m34,805;\s1,1;\s-1,1;\s-1,-1;\s1,-1;\m42,806;\h15;\m42,819;\s1,-2;\s2,-3;\s3,-1;\h2;\s4,1;\s2,3;\s1,2;\U3;\s-1,2;\s-2,2;\s-4,1;\r2;
\s-3,-1;\s-2,-2;\s-1,-2;\D3;\m42,835;\h15;\m46,835;\s-3,3;\s-1,2;\U4;\s1,2;\s3,1;\h11;\m735,795;\D23;\m727,795;\h15;\m752,787;
\s-2,-1;\s-2,-2;\s-1,-4;\D2;\s1,-3;\s2,-2;\s2,-1;\h3;\s2,1;\s3,2;\s1,3;\U2;\s-1,4;\s-3,2;\s-2,1;\r3;\m769,795;\D19;\s1,-3;\s3,-1;
\h2;\m766,787;\h8;\m793,787;\D15;\m793,784;\s-2,2;\s-2,1;\r3;\s-3,-1;\s-2,-2;\s-1,-4;\D2;\s1,-3;\s2,-2;\s3,-1;\h3;\s2,1;\s2,2;
\m802,795;\D23;\m763,607;\s-1,3;\s-2,2;\s-3,1;\r4;\s-2,-1;\s-2,-2;\s-2,-6;\D5;\s2,-6;\s2,-2;\s2,-1;\h4;\s3,1;\s2,2;\s1,3;\m786,607;
\s-1,3;\s-3,2;\s-2,1;\r4;\s-2,-1;\s-2,-2;\s-2,-3;\s-1,-3;\D5;\s1,-3;\s2,-3;\s2,-2;\s2,-1;\h4;\s2,1;\s3,2;\s1,3;\m747,392;\D23;
\m747,392;\s15,-23;\m762,392;\D23;\m786,386;\s-1,2;\s-4,4;\r5;\s-4,-4;\s-1,-2;\s-1,-3;\D5;\s1,-4;\s1,-2;\s2,-2;\s2,-1;\h5;\s2,1;
\s2,2;\s1,2;\m342,636;\s2,2;\s2,1;\h1;\s2,-1;\s1,-1;\s1,-3;\D4;\s-1,-6;\m359,639;\s-1,-3;\s-1,-2;\s-7,-10;\s-2,-4;\s-1,-3;\m365,637;
\h20;\m365,631;\h20;\m393,642;\U1;\s3,3;\s2,1;\h4;\s2,-1;\s1,-1;\s1,-2;\D3;\s-1,-2;\s-2,-3;\s-11,-11;\h16;\m416,626;\s-1,-1;\s1,-1;
\s1,1;\s-1,1;\m428,643;\s2,1;\s3,3;\D23;\put(170,170){\framebox(681,681){}}\m170,170;\U681;\m193,170;\r23;\m182,227;\r12;\m193,284;
\r23;\m182,341;\r12;\m193,397;\r23;\m182,454;\r12;\m193,511;\r23;\m182,568;\r12;\m193,624;\r23;\m182,681;\r12;\m193,738;\r23;
\m182,795;\r12;\m193,851;\r23;\m136,182;\s-4,-1;\s-2,-4;\s-1,-5;\D3;\s1,-6;\s2,-3;\s4,-1;\h2;\s3,1;\s2,3;\s1,6;\U3;\s-1,5;\s-2,4;
\s-3,1;\r2;\m130,290;\U1;\s1,2;\s1,1;\s3,1;\h4;\s2,-1;\s1,-1;\s1,-2;\D2;\s-1,-3;\s-2,-3;\s-11,-11;\h15;\m140,409;\s-11,-16;\h16;
\m140,409;\D23;\m143,519;\s-1,2;\s-3,1;\r2;\s-4,-1;\s-2,-3;\s-1,-6;\D5;\s1,-4;\s2,-3;\s4,-1;\h1;\s3,1;\s2,3;\s1,3;\U1;\s-1,3;\s-2,2;
\s-3,1;\r1;\s-4,-1;\s-2,-2;\s-1,-3;\m135,636;\s-4,-1;\s-1,-3;\D2;\s1,-2;\s2,-1;\s5,-1;\s3,-1;\s2,-2;\s1,-3;\D3;\s-1,-2;\s-1,-1;
\s-3,-1;\r4;\s-4,1;\s-1,1;\s-1,2;\U3;\s1,3;\s2,2;\s8,2;\s2,1;\s1,2;\U2;\s-1,3;\s-3,1;\r4;\m111,745;\s2,1;\s3,3;\D23;\m136,749;
\s-4,-1;\s-2,-3;\s-1,-6;\D3;\s1,-5;\s2,-4;\s4,-1;\h2;\s3,1;\s2,4;\s1,5;\U3;\s-1,6;\s-2,3;\s-3,1;\r2;\m111,858;\s2,1;\s3,4;\D23;
\m130,857;\U1;\s2,4;\s3,1;\h4;\s2,-1;\s2,-4;\D2;\s-1,-2;\s-2,-3;\s-11,-11;\h15;\m170,170;\h681;\m170,193;\D23;\m204,182;\D12;
\m238,182;\D12;\m272,182;\D12;\m306,182;\D12;\m341,193;\D23;\m375,182;\D12;\m409,182;\D12;\m443,182;\D12;\m477,182;\D12;\m511,193;
\D23;\m545,182;\D12;\m579,182;\D12;\m613,182;\D12;\m647,182;\D12;\m681,193;\D23;\m715,182;\D12;\m749,182;\D12;\m783,182;\D12;
\m817,182;\D12;\m851,193;\D23;\m166,143;\s2,1;\s3,4;\D23;\m334,142;\U1;\s1,2;\s3,3;\h5;\s3,-3;\s1,-2;\D2;\s-1,-2;\s-2,-3;\s-11,-11;
\h15;\m505,148;\h12;\s-6,-9;\h3;\s2,-1;\s1,-1;\s1,-4;\D2;\s-1,-3;\s-2,-2;\s-3,-1;\r3;\s-4,1;\s-1,1;\s-1,2;\m684,148;\s-11,-16;\h17;
\m684,148;\D23;\m857,148;\r11;\s-1,-10;\s1,1;\s3,1;\h3;\s4,-1;\s2,-2;\s1,-4;\D2;\s-1,-3;\s-2,-2;\s-4,-1;\r3;\s-3,1;\s-1,1;\s-1,2;
\m851,170;\U681;\m829,170;\h22;\m840,227;\h11;\m829,284;\h22;\m840,341;\h11;\m829,397;\h22;\m840,454;\h11;\m829,511;\h22;\m840,568;
\h11;\m829,624;\h22;\m840,681;\h11;\m829,738;\h22;\m840,795;\h11;\m829,851;\h22;\m170,851;\h681;\m170,829;\U22;\m204,840;\U11;
\m238,840;\U11;\m272,840;\U11;\m306,840;\U11;\m341,829;\U22;\m375,840;\U11;\m409,840;\U11;\m443,840;\U11;\m477,840;\U11;\m511,829;
\U22;\m545,840;\U11;\m579,840;\U11;\m613,840;\U11;\m647,840;\U11;\m681,829;\U22;\m715,840;\U11;\m749,840;\U11;\m783,840;\U11;
\m817,840;\U11;\m851,829;\U22;\m170,225;\s51,21;\s30,13;\s22,9;\s16,7;\s14,6;\s11,5;\s10,4;\s9,3;\s8,4;\s38,16;\s13,7;\s26,12;\s4,2;
\s21,12;\s17,9;\s13,8;\s11,7;\s10,6;\s9,6;\s8,7;\s51,39;\s30,27;\s21,21;\s17,16;\s13,11;\s12,12;\s9,10;\s8,6;\s1,1;\s8,4;\s9,4;
\s28,20;\s7,4;\s7,3;\s12,4;\s18,4;\h7;\s14,-2;\s13,-1;\s4,-1;\s7,-3;\s18,-9;\s10,-4;\s8,-4;\m170,189;\s51,8;\s30,5;\s22,4;\s16,3;
\s14,2;\s11,2;\s10,2;\s9,2;\s8,1;\s23,4;\s28,6;\s30,6;\s21,5;\s17,3;\s13,4;\s11,3;\s10,3;\s9,2;\s8,2;\s12,4;\s16,5;\s23,9;\s22,8;
\s8,3;\s21,9;\s17,7;\s13,6;\s12,5;\s9,5;\s9,4;\s8,4;\s14,6;\s14,5;\s13,5;\s10,3;\s16,4;\s9,3;\s5,1;\s12,2;\s8,1;\h18;\s9,1;\h4;
\s12,-1;\h10;\s5,-1;\s3,-1;\m170,243;\s51,30;\s30,18;\s22,13;\s16,10;\s14,8;\s11,7;\s10,6;\s9,5;\s8,5;\s36,22;\s15,10;\s30,21;
\s21,16;\s17,13;\s13,12;\s11,10;\s10,8;\s9,9;\s8,9;\s34,37;\s17,20;\s20,25;\s10,13;\s21,30;\s17,23;\s13,17;\s12,17;\s9,15;\s8,10;
\s1,1;\s8,7;\s13,13;\s21,23;\s9,8;\s8,7;\s14,8;\s9,5;\s7,2;\s9,1;\h23;\s6,-2;\s10,-3;\s3,-1;\s12,-7;\s10,-5;\s8,-5;\m170,195;\s9,2;
\m188,199;\s6,2;\m194,201;\s3,1;\m205,204;\s9,3;\m223,210;\s8,3;\m240,216;\s7,3;\m247,219;\h1;\m257,222;\s8,4;\m274,229;\s8,4;
\m290,236;\s9,4;\m307,244;\s7,3;\m314,247;\s1,1;\m323,251;\h1;\m324,251;\s8,4;\m340,259;\s1,1;\m341,260;\s7,3;\m356,268;\s8,4;
\m372,276;\s5,3;\m377,279;\s3,1;\m388,285;\s4,2;\m392,287;\s4,2;\m404,294;\s8,4;\m419,302;\s3,2;\m422,304;\s5,3;\m435,312;\s7,5;
\m450,322;\s7,5;\m465,333;\s7,5;\m480,343;\s4,3;\m484,346;\s3,2;\m494,354;\D0;\m494,354;\s8,5;\m509,365;\s2,2;\m511,367;\s5,4;
\m522,377;\s7,6;\m536,388;\s7,6;\m550,400;\s7,6;\m563,412;\s7,7;\m577,425;\s6,6;\m590,438;\s2,2;\m592,440;\s4,4;\m602,451;\s7,6;
\m615,464;\s6,6;\m628,477;\s2,1;\m630,478;\s4,5;\m641,489;\s2,3;\m643,492;\s4,3;\m654,502;\h1;\m655,502;\s6,6;\m667,514;\s6,5;
\m673,519;\s1,1;\m681,526;\D0;\m681,526;\s7,6;\m695,538;\s2,1;\m697,539;\s5,4;\m709,549;\s8,5;\m725,558;\h1;\m726,558;\s6,3;
\m732,561;\s1,1;\m742,564;\s9,2;\m751,566;\D0;\m760,567;\h2;\m762,567;\s7,1;\m778,568;\h5;\m783,568;\s4,-1;\m796,564;\s4,-1;
\m800,563;\s4,-1;\m813,561;\D0;\m813,561;\s9,-3;\m830,554;\s5,-2;\m835,552;\s3,-2;\m170,180;\s9,1;\m188,181;\s9,1;\m206,184;\s9,1;
\m224,186;\s9,1;\m242,188;\s9,1;\m260,191;\s9,1;\m278,194;\s9,1;\m296,197;\s7,1;\m303,198;\h2;\m314,200;\D0;\m314,200;\s9,2;
\m332,203;\s1,1;\m333,204;\s8,1;\m341,205;\D0;\m350,207;\s9,2;\m367,211;\s9,2;\m385,215;\s7,1;\m392,216;\s2,1;\m403,219;\s9,2;
\m421,223;\h1;\m422,223;\s7,2;\m438,227;\s5,2;\m443,229;\s4,1;\m456,232;\s4,1;\m460,233;\s4,1;\m473,237;\s9,2;\m491,242;\s3,1;
\m494,243;\s5,2;\m508,248;\s3,1;\m511,249;\s5,2;\m525,254;\s9,3;\m542,260;\h3;\m545,260;\s6,3;\m559,266;\s3,1;\m562,267;\s6,2;
\m576,273;\s6,2;\m582,275;\s2,1;\m593,280;\s8,4;\m609,287;\s4,2;\m613,289;\s5,2;\m626,294;\s4,2;\m630,296;\s4,2;\m642,302;\s1,1;
\m643,303;\s8,3;\m659,310;\s5,3;\m664,313;\s3,1;\m675,318;\s6,2;\m681,320;\s3,1;\m692,325;\s8,3;\m709,332;\s8,4;\m725,339;\h2;
\m727,339;\s5,2;\m732,341;\s2,1;\m743,344;\s1,1;\m744,345;\s8,1;\m760,348;\s2,1;\m762,349;\s7,1;\m778,352;\h5;\m783,352;\h4;
\m796,353;\h4;\m800,353;\D0;\m800,353;\s5,1;\m814,354;\h9;\m833,353;\h2;\m835,353;\s7,-1;\m170,204;\s9,3;\m187,210;\s8,3;\m195,213;
\s1,1;\m204,217;\s9,4;\m221,224;\U1;\m221,225;\s8,3;\m237,232;\s8,4;\m253,240;\s8,5;\m269,249;\s4,2;\m273,251;\s4,3;\m285,258;\s4,3;
\m289,261;\s4,2;\m301,268;\s2,1;\m303,269;\s5,4;\m316,278;\s7,5;\m331,288;\s2,1;\m333,289;\s5,4;\m346,298;\s7,6;\m360,309;\s8,5;
\m375,320;\s2,1;\m377,321;\s5,4;\m389,331;\s3,2;\m392,333;\s4,3;\m404,342;\s7,6;\m418,353;\s4,4;\m422,357;\s3,2;\m431,365;\s7,6;
\m445,378;\s6,6;\m458,390;\s2,2;\m460,392;\s4,5;\m471,403;\s2,2;\m473,405;\s4,4;\m484,416;\D0;\m484,416;\s6,6;\m496,429;\s7,6;
\m508,442;\s3,3;\m511,445;\s3,4;\m520,456;\s6,7;\m532,470;\s6,7;\m543,484;\s4,5;\m547,489;\s2,2;\m555,498;\s5,8;\m566,513;\s5,7;
\m577,527;\s5,7;\m582,534;\U1;\m587,542;\s5,7;\m592,549;\D0;\m597,557;\s5,8;\m607,572;\s6,8;\m618,587;\s5,8;\m628,602;\s2,2;
\m630,604;\s3,5;\m638,617;\s5,7;\m643,624;\h1;\m649,632;\s5,7;\m659,647;\s5,7;\m664,654;\h1;\m670,661;\s3,5;\m673,666;\s2,3;
\m681,676;\D0;\m681,676;\s6,7;\m692,690;\s6,7;\m704,704;\s3,4;\m707,708;\s3,3;\m716,718;\s2,2;\m718,720;\s4,4;\m729,730;\s3,2;
\m732,732;\s5,3;\m745,739;\s8,4;\m762,746;\D0;\m762,746;\s8,2;\m779,750;\h4;\m783,750;\s5,-1;\m797,747;\s3,-1;\m800,746;\h6;
\m815,745;\s8,-4;\m831,737;\s4,-2;\m835,735;\s4,-3;
\fi\end{picture}
\end{center}
\caption{Shower production functions for muon neutrino (solid line)
and antineutrino (dashed line). Here CC denotes the charged current and NC
the neutral current.}
\end{figure}
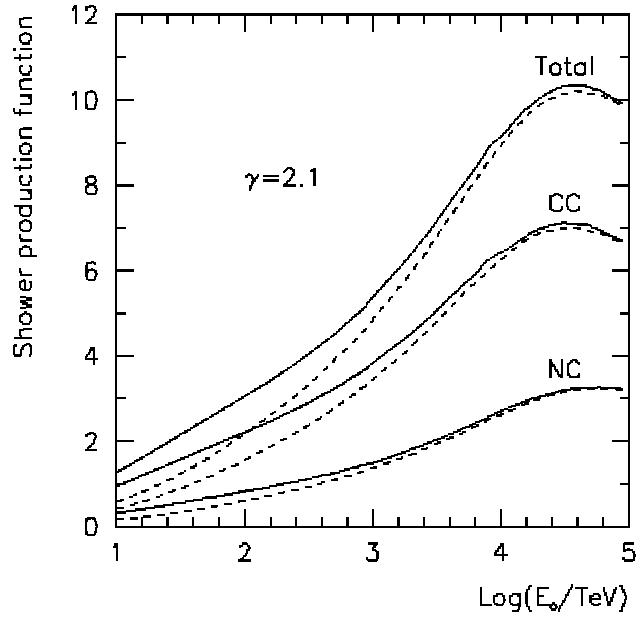
\vspace{0.5cm}

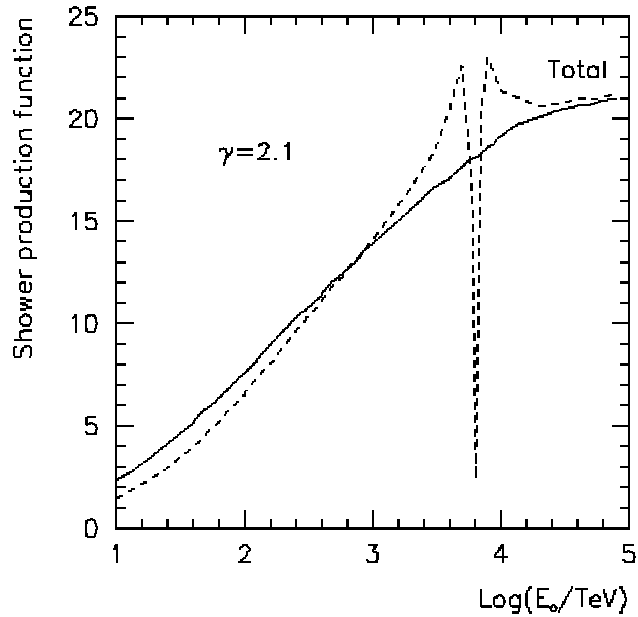
\begin{figure}[b]
\begin{center}
\begin{picture}(800,800)(0,0)\ifx\higzdraft\undefined\newcount\higzdraft\higzdraft=0{}\fi\ifnum\higzdraft>0
\put(0,0){\framebox(800,800){}}\else\ifx\higzstep\undefined\newcount\higzstep\higzstep=0{}\fi\ifnum\higzstep<1\higzstep=2
\fi\ifx\higzxx\undefined\newcount\higzxx\newcount\higzyy\newcount\higzx\newcount\higzy\newcount\higzdx\newcount\higzdy
\newcount\higzlx\newcount\higzly\newcount\higzslope\newcount\higzlen\newcount\higzllen\newcount\higzoffs\newcount\higzloffs
\newcount\higzadash\newcount\higzbdash\newcount\higzcdash\newcount\higzddash\newcount\higzmsize\newcount\higztemp\fi
\def\higzstroke#1,#2,#3,#4;{\advance\higzloffs\higzllen\ifnum\higzloffs>#1\advance\higzloffs-\higzllen\advance\higzloffs-#1
\higzloffs=-\higzloffs\ifnum#2>0\put(\higzlx,\higzly){\line(#3,#4){\higzloffs}}\fi\ifnum#2<0\put(\higzlx,\higzly){\circle*{0}}\fi
\higztemp=\higzloffs\multiply\higztemp#3\advance\higzlx\higztemp\higztemp=\higzloffs\multiply\higztemp#4\advance\higzly\higztemp
\advance\higzllen-\higzloffs\higzloffs=#1\else\ifnum#2>0\put(\higzlx,\higzly){\line(#3,#4){\higzllen}}\fi\ifnum#2<0\put(
\higzlx,\higzly){\circle*{0}}\fi\higzllen=0\fi}\def\higzdashed#1,#2,#3,#4,#5;{{\higzlx=#1\higzly=#2\higzllen=#5\higzloffs=
\higzoffs\loop\ifnum\higzloffs<\higzadash\ifnum\higzadash>1\higzstroke\higzadash,1,#3,#4;\else\higzstroke\higzadash,-1,#3,#4;\fi
\else\ifnum\higzloffs<\higzbdash\higzstroke\higzbdash,0,#3,#4;\else\ifnum\higzloffs<\higzcdash\higztemp=\higzcdash\advance
\higztemp-\higzbdash\ifnum\higztemp>1\higzstroke\higzcdash,1,#3,#4;\else\higzstroke\higzcdash,-1,#3,#4;\fi\else\ifnum\higzloffs<
\higzddash\higzstroke\higzddash,0,#3,#4;\else\higzloffs=0\fi\fi\fi\fi\ifnum\higzllen>0\repeat\global\higzoffs=\higzloffs}}\def
\higzsolid#1,#2,#3,#4,#5;{\put(#1,#2){\line(#3,#4){#5}}}\def\higzhslant#1,#2,#3;{\higzslope=#1\multiply\higzslope1000\advance
\higzslope500\divide\higzslope#2\higzlen=\higzslope\multiply\higzlen\higzstep\divide\higzlen1000\higzdy=0\loop\ifnum
\higzdy<#2\higzx=\higzxx\higzy=\higzyy\higzdx=\higzslope\multiply\higzdx\higzdy\advance\higzdx500\divide\higzdx1000\advance
\higzy\higzdy\multiply\higzdx#3\advance\higzx\higzdx\multiply\higzdx#3\advance\higzdx\higzlen\ifnum\higzdx>#1\advance
\higzlen#1\advance\higzlen-\higzdx\fi\higzline\higzx,\higzy,#3,0,\higzlen;\advance\higzdy\higzstep\repeat}\def\higzvslant#1,#2,#3;{
\higzslope=#2\multiply\higzslope1000\advance\higzslope500\divide\higzslope#1\higzlen=\higzslope\multiply\higzlen\higzstep
\divide\higzlen1000\higzdx=0\loop\ifnum\higzdx<#1\higzx=\higzxx\higzy=\higzyy\higzdy=\higzslope\multiply\higzdy\higzdx\advance
\higzdy500\divide\higzdy1000\advance\higzx\higzdx\multiply\higzdy#3\advance\higzy\higzdy\multiply\higzdy#3\advance
\higzdy\higzlen\ifnum\higzdy>#2\advance\higzlen#2\advance\higzlen-\higzdy\fi\higzline\higzx,\higzy,0,#3,\higzlen;\advance\higzdx
\higzstep\repeat}\def\s#1,#2;{\higzdx=#1{}\ifnum\higzdx<0\higzdx=-\higzdx\fi\higzdy=#2{}\ifnum\higzdy<0\higzdy=-\higzdy\fi\ifnum
\higzdx<\higzdy\ifnum#1<0\advance\higzxx#1\advance\higzyy#2\ifnum#2<0\higzvslant-#1,-#2,1;\else\higzvslant-#1,#2,-1;
\fi\else\ifnum#2<0\higzvslant#1,-#2,-1;\else\higzvslant#1,#2,1;\fi\advance\higzxx#1\advance\higzyy#2\fi\else\ifnum#2<0
\advance\higzxx#1\advance\higzyy#2\ifnum#1<0\higzhslant-#1,-#2,1;\else\higzhslant#1,-#2,-1;\fi\else\ifnum#1<0
\higzhslant-#1,#2,-1;\else\higzhslant#1,#2,1;\fi\advance\higzxx#1\advance\higzyy#2\fi\fi}\def\h#1;{\higzline\higzxx,
\higzyy,1,0,#1;\advance\higzxx#1}\def\r#1;{\higzline\higzxx,\higzyy,-1,0,#1;\advance\higzxx-#1}\def\U#1;{\higzline\higzxx,
\higzyy,0,1,#1;\advance\higzyy#1}\def\D#1;{\higzline\higzxx,\higzyy,0,-1,#1;\advance\higzyy-#1}\def\m#1,#2;{\higzxx=#1
\higzyy=#2}\def\higzdot#1,#2;{\put(#1,#2){\circle*{\higzmsize}}}\def\higzplus#1,#2;{\higzx=#1\multiply\higzx2\advance\higzx-
\higzmsize\divide\higzx2\put(\higzx,#2){\line(1,0){\higzmsize}}\higzy=#2\multiply\higzy2\advance\higzy-\higzmsize\divide
\higzy2\put(#1,\higzy){\line(0,1){\higzmsize}}}\def\higzstar#1,#2;{\higzplus#1,#2;\higzcross#1,#2;}\def
\higzcircle#1,#2;{\put(#1,#2){\circle{\higzmsize}}}\def\higzcross#1,#2;{\let\higzsave\higzline\let\higzline\higzsolid\higzlx=#1
\multiply\higzlx2\advance\higzlx-\higzmsize\divide\higzlx2\higzly=#2\multiply\higzly2\advance\higzly-\higzmsize\divide
\higzly2\m\higzlx,\higzly;\s\higzmsize,\higzmsize;\higzly=#2\multiply\higzly2\advance\higzly\higzmsize\divide\higzly2\m\higzlx,
\higzly;\s\higzmsize,-\higzmsize;\let\higzline\higzsave}\def\p#1,#2;{\higzmarker#1,#2;}\def\f#1,#2;{\put(\higzxx,
\higzyy){\makebox(#1,#2)[lb]{\rule{#1\unitlength}{#2\unitlength}}}}
\let\higzline=\higzsolid\let\higzline=\higzsolid\m655,89;\D23;\m655,66;\h13;\m678,81;\s-2,-1;\s-3,-2;\s-1,-4;\D2;\s1,-3;\s3,-2;
\s2,-1;\h3;\s2,1;\s2,2;\s1,3;\U2;\s-1,4;\s-2,2;\s-2,1;\r3;\m706,81;\D17;\s-1,-4;\s-1,-1;\s-2,-1;\r4;\s-2,1;\m706,78;\s-2,2;\s-2,1;
\r4;\s-2,-1;\s-2,-2;\s-1,-4;\D2;\s1,-3;\s2,-2;\s2,-1;\h4;\s2,1;\s2,2;\m722,93;\s-2,-2;\s-4,-8;\s-1,-5;\D5;\s1,-5;\s4,-8;\s2,-2;
\m730,89;\D23;\m730,89;\h14;\m730,78;\h8;\m730,66;\h14;\m750,68;\s-2,-2;\s-1,-2;\D1;\s3,-3;\h2;\s3,3;\U1;\s-1,2;\s-2,2;\r2;\m778,93;
\s-20,-35;\m789,89;\D23;\m781,89;\h15;\m800,74;\h13;\U3;\s-1,2;\s-1,1;\s-2,1;\r3;\s-2,-1;\s-2,-2;\s-2,-4;\D2;\s2,-3;\s2,-2;\s2,-1;
\h3;\s2,1;\s2,2;\m818,89;\s8,-23;\m835,89;\s-9,-23;\m839,93;\s3,-2;\s4,-8;\s1,-5;\D5;\s-1,-5;\s-4,-8;\s-3,-2;\m37,413;\s-2,-2;
\s-1,-3;\D4;\s1,-4;\s2,-2;\h2;\s3,1;\s1,1;\s3,9;\s1,2;\s1,1;\s2,1;\h4;\s2,-2;\s1,-3;\D4;\s-1,-4;\s-2,-2;\m34,421;\h23;\m46,421;
\s-3,3;\s-1,2;\U4;\s1,2;\s3,1;\h11;\m42,446;\s1,-2;\s2,-2;\s3,-2;\h2;\s4,2;\s2,2;\s1,2;\U3;\s-1,2;\s-2,2;\s-4,2;\r2;\s-3,-2;\s-2,-2;
\s-1,-2;\D3;\m42,461;\s15,4;\m42,470;\s15,-5;\m42,470;\s15,4;\m42,478;\s15,-4;\m48,485;\U13;\r2;\s-2,-1;\s-1,-1;\s-1,-3;\D3;\s1,-2;
\s2,-2;\s3,-1;\h2;\s4,1;\s2,2;\s1,2;\U3;\s-1,3;\s-2,2;\m42,505;\h15;\m48,505;\s-3,1;\s-2,3;\s-1,2;\U3;\m42,530;\h22;\m45,530;\s-2,2;
\s-1,3;\U3;\s1,2;\s2,2;\s3,1;\h2;\s4,-1;\s2,-2;\s1,-2;\D3;\s-1,-3;\s-2,-2;\m42,551;\h15;\m48,551;\s-3,1;\s-2,2;\s-1,2;\U3;\m42,569;
\s1,-2;\s2,-2;\s3,-1;\h2;\s4,1;\s2,2;\s1,2;\U3;\s-1,3;\s-2,2;\s-4,1;\r2;\s-3,-1;\s-2,-2;\s-1,-3;\D3;\m34,597;\h23;\m45,597;\s-2,-2;
\s-1,-2;\D3;\s1,-3;\s2,-2;\s3,-1;\h2;\s4,1;\s2,2;\s1,3;\U3;\s-1,2;\s-2,2;\m42,606;\h10;\s4,1;\s1,2;\U3;\s-1,3;\s-4,3;\m42,618;\h15;
\m45,638;\s-2,-2;\s-1,-2;\D3;\s1,-2;\s2,-3;\s3,-1;\h2;\s4,1;\s2,3;\s1,2;\U3;\s-1,2;\s-2,2;\m34,647;\h18;\s4,1;\s1,2;\U2;\m42,644;
\U7;\m34,658;\s1,1;\s-1,1;\s-1,-1;\s1,-1;\m42,659;\h15;\m42,672;\s1,-2;\s2,-3;\s3,-1;\h2;\s4,1;\s2,3;\s1,2;\U3;\s-1,2;\s-2,2;\s-4,1;
\r2;\s-3,-1;\s-2,-2;\s-1,-2;\D3;\m42,688;\h15;\m46,688;\s-3,3;\s-1,2;\U4;\s1,2;\s3,1;\h11;\m34,726;\D2;\s1,-2;\s3,-2;\h19;\m42,717;
\U8;\m42,732;\h10;\s4,1;\s1,3;\U3;\s-1,2;\s-4,3;\m42,744;\h15;\m42,753;\h15;\m46,753;\s-3,3;\s-1,2;\U4;\s1,2;\s3,1;\h11;\m45,785;
\s-2,-2;\s-1,-2;\D3;\s1,-2;\s2,-3;\s3,-1;\h2;\s4,1;\s2,3;\s1,2;\U3;\s-1,2;\s-2,2;\m34,794;\h18;\s4,1;\s1,2;\U2;\m42,791;\U7;
\m34,805;\s1,1;\s-1,1;\s-1,-1;\s1,-1;\m42,806;\h15;\m42,819;\s1,-2;\s2,-3;\s3,-1;\h2;\s4,1;\s2,3;\s1,2;\U3;\s-1,2;\s-2,2;\s-4,1;\r2;
\s-3,-1;\s-2,-2;\s-1,-2;\D3;\m42,835;\h15;\m46,835;\s-3,3;\s-1,2;\U4;\s1,2;\s3,1;\h11;\m308,672;\s2,3;\s2,1;\h1;\s2,-1;\s1,-1;
\s1,-4;\D4;\s-1,-5;\m325,676;\s-1,-4;\s-1,-2;\s-7,-9;\s-2,-5;\s-1,-3;\m331,674;\h20;\m331,667;\h20;\m359,678;\U1;\s1,2;\s4,2;\h4;
\s2,-1;\s1,-1;\s1,-2;\D2;\s-3,-6;\s-11,-10;\h15;\m382,663;\s-1,-1;\s1,-1;\s1,1;\s-1,1;\m394,679;\s2,1;\s3,3;\D22;\m752,792;\D22;
\m744,792;\h15;\m769,785;\s-2,-1;\s-2,-3;\s-1,-3;\D2;\s1,-3;\s2,-2;\s2,-1;\h3;\s3,1;\s2,2;\s1,3;\U2;\s-1,3;\s-2,3;\s-3,1;\r3;
\m786,792;\D18;\s1,-3;\s3,-1;\h2;\m783,785;\h8;\m810,785;\D15;\m810,781;\s-2,3;\s-2,1;\r3;\s-3,-1;\s-2,-3;\s-1,-3;\D2;\s1,-3;\s2,-2;
\s3,-1;\h3;\s2,1;\s2,2;\m819,792;\D22;\put(170,170){\framebox(681,681){}}\m170,170;\U681;\m193,170;\r23;\m182,197;\r12;\m182,225;
\r12;\m182,252;\r12;\m182,279;\r12;\m193,306;\r23;\m182,334;\r12;\m182,361;\r12;\m182,388;\r12;\m182,415;\r12;\m193,443;\r23;
\m182,470;\r12;\m182,497;\r12;\m182,524;\r12;\m182,552;\r12;\m193,579;\r23;\m182,606;\r12;\m182,633;\r12;\m182,661;\r12;\m182,688;
\r12;\m193,715;\r23;\m182,742;\r12;\m182,770;\r12;\m182,797;\r12;\m182,824;\r12;\m193,851;\r23;\m136,182;\s-4,-1;\s-2,-4;\s-1,-5;
\D3;\s1,-6;\s2,-3;\s4,-1;\h2;\s3,1;\s2,3;\s1,6;\U3;\s-1,5;\s-2,4;\s-3,1;\r2;\m142,318;\r11;\s-1,-10;\s1,1;\s4,1;\h3;\s3,-1;\s2,-2;
\s1,-3;\D2;\s-1,-4;\s-2,-2;\s-3,-1;\r3;\s-4,1;\s-1,1;\s-1,2;\m111,450;\s2,1;\s3,3;\D23;\m136,454;\s-4,-1;\s-2,-3;\s-1,-6;\D3;\s1,-5;
\s2,-4;\s4,-1;\h2;\s3,1;\s2,4;\s1,5;\U3;\s-1,6;\s-2,3;\s-3,1;\r2;\m111,586;\s2,1;\s3,3;\D22;\m142,590;\r11;\s-1,-10;\s1,2;\s4,1;\h3;
\s3,-1;\s2,-3;\s1,-3;\D2;\s-1,-3;\s-2,-2;\s-3,-1;\r3;\s-4,1;\s-1,1;\s-1,2;\m109,721;\U1;\s1,2;\s1,1;\s2,1;\h4;\s4,-2;\s1,-2;\D2;
\s-1,-2;\s-3,-3;\s-10,-11;\h15;\m136,726;\s-4,-1;\s-2,-3;\s-1,-5;\D4;\s1,-5;\s2,-3;\s4,-1;\h2;\s3,1;\s2,3;\s1,5;\U4;\s-1,5;\s-2,3;
\s-3,1;\r2;\m109,857;\U1;\s2,4;\s2,1;\h4;\s2,-1;\s2,-2;\s1,-2;\D2;\s-1,-2;\s-3,-3;\s-10,-11;\h15;\m142,863;\r11;\s-1,-10;\s1,1;
\s4,1;\h3;\s3,-1;\s2,-2;\s1,-3;\D3;\s-1,-3;\s-2,-2;\s-3,-1;\r3;\s-4,1;\s-1,1;\s-1,2;\m170,170;\h681;\m170,193;\D23;\m204,182;\D12;
\m238,182;\D12;\m272,182;\D12;\m306,182;\D12;\m341,193;\D23;\m375,182;\D12;\m409,182;\D12;\m443,182;\D12;\m477,182;\D12;\m511,193;
\D23;\m545,182;\D12;\m579,182;\D12;\m613,182;\D12;\m647,182;\D12;\m681,193;\D23;\m715,182;\D12;\m749,182;\D12;\m783,182;\D12;
\m817,182;\D12;\m851,193;\D23;\m166,143;\s2,1;\s3,4;\D23;\m334,142;\U1;\s1,2;\s3,3;\h5;\s3,-3;\s1,-2;\D2;\s-1,-2;\s-2,-3;\s-11,-11;
\h15;\m505,148;\h12;\s-6,-9;\h3;\s2,-1;\s1,-1;\s1,-4;\D2;\s-1,-3;\s-2,-2;\s-3,-1;\r3;\s-4,1;\s-1,1;\s-1,2;\m684,148;\s-11,-16;\h17;
\m684,148;\D23;\m857,148;\r11;\s-1,-10;\s1,1;\s3,1;\h3;\s4,-1;\s2,-2;\s1,-4;\D2;\s-1,-3;\s-2,-2;\s-4,-1;\r3;\s-3,1;\s-1,1;\s-1,2;
\m851,170;\U681;\m829,170;\h22;\m840,197;\h11;\m840,225;\h11;\m840,252;\h11;\m840,279;\h11;\m829,306;\h22;\m840,334;\h11;\m840,361;
\h11;\m840,388;\h11;\m840,415;\h11;\m829,443;\h22;\m840,470;\h11;\m840,497;\h11;\m840,524;\h11;\m840,552;\h11;\m829,579;\h22;
\m840,606;\h11;\m840,633;\h11;\m840,661;\h11;\m840,688;\h11;\m829,715;\h22;\m840,742;\h11;\m840,770;\h11;\m840,797;\h11;\m840,824;
\h11;\m829,851;\h22;\m170,851;\h681;\m170,829;\U22;\m204,840;\U11;\m238,840;\U11;\m272,840;\U11;\m306,840;\U11;\m341,829;\U22;
\m375,840;\U11;\m409,840;\U11;\m443,840;\U11;\m477,840;\U11;\m511,829;\U22;\m545,840;\U11;\m579,840;\U11;\m613,840;\U11;\m647,840;
\U11;\m681,829;\U22;\m715,840;\U11;\m749,840;\U11;\m783,840;\U11;\m817,840;\U11;\m851,829;\U22;\m170,235;\s22,13;\s22,16;\s7,6;
\s30,24;\s22,18;\s16,17;\s14,12;\s30,30;\s8,7;\s16,16;\s35,38;\s25,26;\s5,4;\s21,19;\s6,7;\s9,11;\s2,2;\s13,11;\s11,10;\s10,10;
\s9,11;\s8,7;\s51,46;\s14,13;\s11,11;\s5,4;\s21,13;\s10,9;\s7,6;\s9,9;\s4,2;\s5,1;\s4,3;\s3,3;\s9,8;\s9,5;\s8,9;\s7,6;\s8,6;\s9,5;
\s27,9;\s19,7;\s11,3;\s17,4;\s4,1;\s17,1;\s13,5;\s10,2;\h12;\m170,211;\D0;\m170,211;\s8,4;\m186,220;\s8,4;\m202,229;\s8,4;\m218,238;
\s3,2;\m221,240;\s4,3;\m233,248;\s7,6;\m247,259;\s4,3;\m251,262;\s3,3;\m261,270;\s7,6;\m275,282;\s7,6;\m289,294;\D0;\m289,294;\s6,7;
\m301,307;\s2,2;\m303,309;\s5,5;\m314,320;\s6,7;\m326,334;\s6,6;\m338,347;\s3,2;\m341,349;\s3,5;\m350,361;\s6,7;\m362,375;\s6,7;
\m374,389;\s6,6;\m385,402;\s6,7;\m397,416;\s6,8;\m408,431;\s6,7;\m419,445;\s3,3;\m422,448;\s3,4;\m431,459;\s6,6;\m443,472;\s6,8;
\m454,487;\s6,7;\m460,494;\D0;\m466,501;\s6,6;\m478,515;\s5,7;\m490,528;\s4,5;\m494,533;\s2,2;\m501,542;\s2,2;\m503,544;\s4,5;
\m513,556;\s6,7;\m524,570;\s5,8;\m535,585;\s5,7;\m545,600;\s6,7;\m556,614;\s6,8;\m566,629;\s5,8;\m576,644;\s5,8;\m586,660;\s5,7;
\m595,676;\s3,8;\m601,693;\s3,8;\m607,710;\s4,8;\m614,727;\s2,8;\m619,744;\s2,9;\m623,762;\s3,8;\m628,779;\s2,6;\m630,785;\D2;
\m631,774;\D9;\m632,755;\s1,-9;\m634,737;\D9;\m635,719;\s1,-9;\m637,701;\D9;\m638,683;\s1,-9;\m640,665;\D9;\m641,647;\s1,-9;
\m642,629;\s1,-9;\m643,620;\D0;\m643,611;\D9;\m644,592;\D9;\m644,574;\D9;\m644,556;\D9;\m644,538;\D9;\m644,520;\s1,-9;\m645,502;\D9;
\m645,484;\D10;\m645,465;\D9;\m645,447;\D9;\m646,429;\D9;\m646,411;\D9;\m646,393;\D9;\m646,375;\D10;\m646,356;\s1,-9;\m647,338;\D9;
\m647,320;\D9;\m647,302;\D9;\m647,284;\D9;\m648,266;\D9;\m648,247;\D9;\m648,243;\U9;\m648,261;\U9;\m649,279;\U9;\m649,297;\U9;
\m649,315;\U9;\m649,334;\U9;\m650,352;\U9;\m650,370;\U9;\m650,388;\U9;\m650,406;\U9;\m651,424;\U9;\m651,442;\U10;\m651,461;\U9;
\m651,479;\U9;\m652,497;\U9;\m652,515;\U9;\m652,533;\U9;\m652,551;\U9;\m653,570;\U9;\m653,588;\U9;\m653,606;\U9;\m653,624;\U9;
\m654,642;\U9;\m654,660;\U9;\m654,679;\U9;\m654,697;\U9;\m655,715;\U6;\m655,721;\U3;\m656,733;\s1,9;\m659,751;\s1,9;\m661,769;\s1,9;
\m663,787;\s1,9;\m668,788;\s3,-9;\m675,771;\s3,-9;\m681,754;\D1;\m681,753;\s8,-3;\m697,746;\s9,-3;\m714,740;\s8,-4;\m731,733;\s1,-1;
\m732,732;\s8,1;\m749,734;\s9,1;\m767,737;\s8,3;\m784,743;\s9,-1;\m802,741;\s8,2;\m819,745;\s6,2;\m825,747;\h3;
\fi\end{picture}
\end{center}
\caption{Shower production functions for electron neutrino (solid
line) and antineutrino (dashed line).}
\end{figure}
\clearpage

\begin{figure}[t]
\begin{center}
\begin{picture}(800,800)(0,0)\ifx\higzdraft\undefined\newcount\higzdraft\higzdraft=0{}\fi\ifnum\higzdraft>0
\put(0,0){\framebox(800,800){}}\else\ifx\higzstep\undefined\newcount\higzstep\higzstep=0{}\fi\ifnum\higzstep<1\higzstep=2
\fi\ifx\higzxx\undefined\newcount\higzxx\newcount\higzyy\newcount\higzx\newcount\higzy\newcount\higzdx\newcount\higzdy
\newcount\higzlx\newcount\higzly\newcount\higzslope\newcount\higzlen\newcount\higzllen\newcount\higzoffs\newcount\higzloffs
\newcount\higzadash\newcount\higzbdash\newcount\higzcdash\newcount\higzddash\newcount\higzmsize\newcount\higztemp\fi
\def\higzstroke#1,#2,#3,#4;{\advance\higzloffs\higzllen\ifnum\higzloffs>#1\advance\higzloffs-\higzllen\advance\higzloffs-#1
\higzloffs=-\higzloffs\ifnum#2>0\put(\higzlx,\higzly){\line(#3,#4){\higzloffs}}\fi\ifnum#2<0\put(\higzlx,\higzly){\circle*{0}}\fi
\higztemp=\higzloffs\multiply\higztemp#3\advance\higzlx\higztemp\higztemp=\higzloffs\multiply\higztemp#4\advance\higzly\higztemp
\advance\higzllen-\higzloffs\higzloffs=#1\else\ifnum#2>0\put(\higzlx,\higzly){\line(#3,#4){\higzllen}}\fi\ifnum#2<0\put(
\higzlx,\higzly){\circle*{0}}\fi\higzllen=0\fi}\def\higzdashed#1,#2,#3,#4,#5;{{\higzlx=#1\higzly=#2\higzllen=#5\higzloffs=
\higzoffs\loop\ifnum\higzloffs<\higzadash\ifnum\higzadash>1\higzstroke\higzadash,1,#3,#4;\else\higzstroke\higzadash,-1,#3,#4;\fi
\else\ifnum\higzloffs<\higzbdash\higzstroke\higzbdash,0,#3,#4;\else\ifnum\higzloffs<\higzcdash\higztemp=\higzcdash\advance
\higztemp-\higzbdash\ifnum\higztemp>1\higzstroke\higzcdash,1,#3,#4;\else\higzstroke\higzcdash,-1,#3,#4;\fi\else\ifnum\higzloffs<
\higzddash\higzstroke\higzddash,0,#3,#4;\else\higzloffs=0\fi\fi\fi\fi\ifnum\higzllen>0\repeat\global\higzoffs=\higzloffs}}\def
\higzsolid#1,#2,#3,#4,#5;{\put(#1,#2){\line(#3,#4){#5}}}\def\higzhslant#1,#2,#3;{\higzslope=#1\multiply\higzslope1000\advance
\higzslope500\divide\higzslope#2\higzlen=\higzslope\multiply\higzlen\higzstep\divide\higzlen1000\higzdy=0\loop\ifnum
\higzdy<#2\higzx=\higzxx\higzy=\higzyy\higzdx=\higzslope\multiply\higzdx\higzdy\advance\higzdx500\divide\higzdx1000\advance
\higzy\higzdy\multiply\higzdx#3\advance\higzx\higzdx\multiply\higzdx#3\advance\higzdx\higzlen\ifnum\higzdx>#1\advance
\higzlen#1\advance\higzlen-\higzdx\fi\higzline\higzx,\higzy,#3,0,\higzlen;\advance\higzdy\higzstep\repeat}\def\higzvslant#1,#2,#3;{
\higzslope=#2\multiply\higzslope1000\advance\higzslope500\divide\higzslope#1\higzlen=\higzslope\multiply\higzlen\higzstep
\divide\higzlen1000\higzdx=0\loop\ifnum\higzdx<#1\higzx=\higzxx\higzy=\higzyy\higzdy=\higzslope\multiply\higzdy\higzdx\advance
\higzdy500\divide\higzdy1000\advance\higzx\higzdx\multiply\higzdy#3\advance\higzy\higzdy\multiply\higzdy#3\advance
\higzdy\higzlen\ifnum\higzdy>#2\advance\higzlen#2\advance\higzlen-\higzdy\fi\higzline\higzx,\higzy,0,#3,\higzlen;\advance\higzdx
\higzstep\repeat}\def\s#1,#2;{\higzdx=#1{}\ifnum\higzdx<0\higzdx=-\higzdx\fi\higzdy=#2{}\ifnum\higzdy<0\higzdy=-\higzdy\fi\ifnum
\higzdx<\higzdy\ifnum#1<0\advance\higzxx#1\advance\higzyy#2\ifnum#2<0\higzvslant-#1,-#2,1;\else\higzvslant-#1,#2,-1;
\fi\else\ifnum#2<0\higzvslant#1,-#2,-1;\else\higzvslant#1,#2,1;\fi\advance\higzxx#1\advance\higzyy#2\fi\else\ifnum#2<0
\advance\higzxx#1\advance\higzyy#2\ifnum#1<0\higzhslant-#1,-#2,1;\else\higzhslant#1,-#2,-1;\fi\else\ifnum#1<0
\higzhslant-#1,#2,-1;\else\higzhslant#1,#2,1;\fi\advance\higzxx#1\advance\higzyy#2\fi\fi}\def\h#1;{\higzline\higzxx,
\higzyy,1,0,#1;\advance\higzxx#1}\def\r#1;{\higzline\higzxx,\higzyy,-1,0,#1;\advance\higzxx-#1}\def\U#1;{\higzline\higzxx,
\higzyy,0,1,#1;\advance\higzyy#1}\def\D#1;{\higzline\higzxx,\higzyy,0,-1,#1;\advance\higzyy-#1}\def\m#1,#2;{\higzxx=#1
\higzyy=#2}\def\higzdot#1,#2;{\put(#1,#2){\circle*{\higzmsize}}}\def\higzplus#1,#2;{\higzx=#1\multiply\higzx2\advance\higzx-
\higzmsize\divide\higzx2\put(\higzx,#2){\line(1,0){\higzmsize}}\higzy=#2\multiply\higzy2\advance\higzy-\higzmsize\divide
\higzy2\put(#1,\higzy){\line(0,1){\higzmsize}}}\def\higzstar#1,#2;{\higzplus#1,#2;\higzcross#1,#2;}\def
\higzcircle#1,#2;{\put(#1,#2){\circle{\higzmsize}}}\def\higzcross#1,#2;{\let\higzsave\higzline\let\higzline\higzsolid\higzlx=#1
\multiply\higzlx2\advance\higzlx-\higzmsize\divide\higzlx2\higzly=#2\multiply\higzly2\advance\higzly-\higzmsize\divide
\higzly2\m\higzlx,\higzly;\s\higzmsize,\higzmsize;\higzly=#2\multiply\higzly2\advance\higzly\higzmsize\divide\higzly2\m\higzlx,
\higzly;\s\higzmsize,-\higzmsize;\let\higzline\higzsave}\def\p#1,#2;{\higzmarker#1,#2;}\def\f#1,#2;{\put(\higzxx,
\higzyy){\makebox(#1,#2)[lb]{\rule{#1\unitlength}{#2\unitlength}}}}
\let\higzline=\higzsolid\m713,85;\s-2,2;\s-4,2;\r4;\s-3,-2;\s-2,-2;\D2;\s1,-2;\s1,-1;\s2,-1;\s7,-2;\s2,-1;\s2,-4;\D3;\s-2,-2;
\s-4,-1;\r4;\s-3,1;\s-2,2;\m738,93;\s-20,-35;\m744,89;\D23;\m744,89;\s15,-23;\m759,89;\D23;\m779,81;\D15;\m779,74;\s1,4;\s2,2;\s2,1;
\h3;\m805,81;\D15;\m805,78;\s-2,2;\s-3,1;\r3;\s-2,-1;\s-2,-2;\s-1,-4;\D2;\s1,-3;\s2,-2;\s2,-1;\h3;\s3,1;\s2,2;\m814,89;\D19;\s2,-3;
\s2,-1;\h2;\m811,81;\h8;\m825,89;\s1,-2;\s1,2;\s-1,1;\s-1,-1;\m826,81;\D15;\m839,81;\s-2,-1;\s-2,-2;\s-1,-4;\D2;\s1,-3;\s2,-2;
\s2,-1;\h4;\s2,1;\s2,2;\s1,3;\U2;\s-1,4;\s-2,2;\s-2,1;\r4;\m34,475;\h23;\m57,475;\U13;\m42,497;\s1,-2;\s2,-2;\s3,-1;\h2;\s4,1;\s2,2;
\s1,2;\U4;\s-1,2;\s-2,2;\s-4,1;\r2;\s-3,-1;\s-2,-2;\s-1,-2;\D4;\m42,525;\h17;\s3,-1;\s1,-1;\s1,-2;\D3;\s-1,-2;\m45,525;\s-2,-2;
\s-1,-2;\D3;\s1,-2;\s2,-2;\s3,-1;\h2;\s4,1;\s2,2;\s1,2;\U3;\s-1,2;\s-2,2;\m34,533;\h23;\m34,533;\h23;\m34,533;\U5;\m57,533;\U5;
\m34,544;\h23;\m34,544;\U10;\s1,3;\s1,1;\s2,1;\h3;\s2,-1;\s1,-1;\s1,-3;\D10;\m45,552;\s12,7;\m42,579;\h15;\m45,579;\s-2,-3;\s-1,-2;
\D3;\s1,-2;\s2,-2;\s3,-1;\h2;\s4,1;\s2,2;\s1,2;\U3;\s-1,2;\s-2,3;\m34,588;\h18;\s4,1;\s1,3;\U2;\m42,585;\U8;\m48,599;\U13;\r2;
\s-2,-1;\s-1,-1;\s-1,-2;\D3;\s1,-3;\s2,-2;\s3,-1;\h2;\s4,1;\s2,2;\s1,3;\U3;\s-1,2;\s-2,2;\m30,637;\s34,-19;\m30,651;\s2,-2;\s3,-2;
\s4,-2;\s6,-1;\h4;\s6,1;\s4,2;\s3,2;\s2,2;\m42,656;\s15,7;\m42,669;\s15,-6;\s4,-2;\s2,-2;\s1,-3;\D1;\m48,675;\U13;\r2;\s-2,-1;
\s-1,-1;\s-1,-3;\D3;\s1,-2;\s2,-2;\s3,-1;\h2;\s4,1;\s2,2;\s1,2;\U3;\s-1,3;\s-2,2;\m42,707;\h15;\m45,707;\s-2,-2;\s-1,-2;\D3;\s1,-2;
\s2,-3;\s3,-1;\h2;\s4,1;\s2,3;\s1,2;\U3;\s-1,2;\s-2,2;\m42,716;\h15;\m48,716;\s-3,1;\s-2,2;\s-1,2;\U4;\m35,728;\U10;\m31,743;\s-3,3;
\h12;\m34,755;\h23;\m42,766;\s10,-11;\m48,759;\s9,8;\m42,773;\h15;\m46,773;\s-3,3;\s-1,3;\U3;\s1,2;\s3,1;\h11;\m46,785;\s-3,3;
\s-1,3;\U3;\s1,2;\s3,1;\h11;\m35,803;\U10;\m31,818;\D0;\r2;\U1;\s-1,1;\U2;\s1,1;\s2,1;\h1;\s2,-2;\s6,-5;\U8;\m30,829;\s2,3;\s3,2;
\s4,2;\s6,1;\h4;\s6,-1;\s4,-2;\s3,-2;\s2,-3;\m34,848;\h23;\m34,848;\h23;\m34,843;\U5;\m57,843;\U5;\m567,649;\s-2,3;\s-3,1;\r5;
\s-3,-1;\s-2,-3;\D2;\s1,-2;\s1,-1;\s2,-1;\s7,-2;\s2,-1;\s1,-1;\s1,-3;\D3;\s-2,-2;\s-3,-1;\r5;\s-3,1;\s-2,2;\m575,653;\D23;\m575,653;
\h9;\s3,-1;\s2,-2;\s1,-2;\D3;\s-1,-2;\s-2,-1;\s-3,-1;\r9;\m597,645;\D15;\m597,641;\s3,3;\s3,1;\h3;\s2,-1;\s1,-3;\D11;\m609,641;
\s3,3;\s3,1;\h3;\s2,-1;\s1,-3;\D11;\m642,645;\D15;\m642,642;\s-3,2;\s-2,1;\r3;\s-2,-1;\s-2,-2;\s-1,-3;\D3;\s1,-3;\s2,-2;\s2,-1;\h3;
\s2,1;\s3,2;\m649,645;\s12,-15;\m661,645;\s-12,-15;\m491,513;\s-2,2;\s-3,1;\r4;\s-4,-1;\s-2,-2;\D2;\s1,-2;\s1,-1;\s2,-1;\s7,-2;
\s3,-3;\s1,-2;\D3;\s-2,-2;\s-3,-1;\r4;\s-4,1;\s-2,2;\m499,516;\D22;\m499,516;\h10;\s3,-1;\s1,-1;\s1,-2;\D3;\s-1,-2;\s-1,-1;\s-3,-1;
\r10;\m522,509;\D15;\m522,505;\s3,3;\s2,1;\h3;\s2,-1;\s1,-3;\D11;\m533,505;\s4,3;\s2,1;\h3;\s2,-1;\s1,-3;\D11;\m553,516;\s1,-1;
\s1,1;\s-1,2;\s-1,-2;\m554,509;\D15;\m563,509;\D15;\m563,505;\s3,3;\s2,1;\h3;\s2,-1;\s2,-3;\D11;\m264,513;\s-2,2;\s-3,1;\r4;\s-4,-1;
\s-2,-2;\D2;\s1,-2;\s1,-1;\s2,-1;\s7,-2;\s3,-3;\s1,-2;\D3;\s-2,-2;\s-3,-1;\r4;\s-4,1;\s-2,2;\m272,516;\D22;\m272,516;\h7;\s4,-1;
\s2,-2;\s1,-2;\s1,-3;\D6;\s-1,-3;\s-1,-2;\s-2,-2;\s-4,-1;\r7;\m309,513;\s-3,2;\s-3,1;\r4;\s-3,-1;\s-3,-2;\D2;\s3,-3;\s2,-1;\s6,-2;
\s2,-2;\s2,-1;\s1,-2;\D3;\s-3,-2;\s-3,-1;\r4;\s-3,1;\s-3,2;\m330,513;\s-2,2;\s-3,1;\r4;\s-4,-1;\s-2,-2;\D2;\s1,-2;\s1,-1;\s2,-1;
\s7,-2;\s3,-3;\s1,-2;\D3;\s-2,-2;\s-3,-1;\r4;\s-4,1;\s-2,2;\m629,380;\D22;\m629,380;\h9;\s4,-1;\s1,-1;\s1,-2;\D2;\s-2,-4;\s-4,-1;
\m629,369;\h9;\s4,-1;\s1,-1;\s1,-2;\D3;\s-1,-2;\s-1,-1;\s-4,-1;\r9;\m650,380;\s1,-1;\s1,1;\s-1,1;\s-1,-1;\m651,373;\D15;\m340,292;
\s-2,2;\s-3,1;\r5;\s-3,-1;\s-2,-2;\D2;\s2,-4;\s2,-1;\s7,-2;\s2,-1;\s1,-1;\s1,-2;\D3;\s-2,-3;\s-3,-1;\r5;\s-3,1;\s-2,3;\m348,295;
\D23;\m348,295;\h9;\s3,-1;\s2,-1;\s1,-2;\D2;\s-1,-3;\s-2,-1;\s-3,-1;\m348,284;\h9;\s3,-1;\s2,-1;\s1,-2;\D3;\s-1,-2;\s-2,-2;\s-3,-1;
\r9;\put(170,170){\framebox(681,681){}}\m170,170;\U681;\m193,170;\r23;\m182,204;\r12;\m182,238;\r12;\m182,272;\r12;\m182,306;\r12;
\m193,341;\r23;\m182,375;\r12;\m182,409;\r12;\m182,443;\r12;\m182,477;\r12;\m193,511;\r23;\m182,545;\r12;\m182,579;\r12;\m182,613;
\r12;\m182,647;\r12;\m193,681;\r23;\m182,715;\r12;\m182,749;\r12;\m182,783;\r12;\m182,817;\r12;\m193,851;\r23;\m136,182;\s-4,-1;
\s-2,-4;\s-1,-5;\D3;\s1,-6;\s2,-3;\s4,-1;\h2;\s3,1;\s2,3;\s1,6;\U3;\s-1,5;\s-2,4;\s-3,1;\r2;\m132,348;\s3,1;\s3,3;\D23;\m130,517;
\U1;\s1,2;\s1,1;\s3,1;\h4;\s2,-1;\s1,-1;\s1,-2;\D2;\s-1,-3;\s-2,-3;\s-11,-11;\h15;\m131,692;\h12;\s-6,-8;\h3;\s2,-1;\s1,-1;\s1,-4;
\D2;\s-1,-3;\s-2,-2;\s-3,-1;\r3;\s-4,1;\s-1,1;\s-1,2;\m140,863;\s-11,-16;\h16;\m140,863;\D23;\m170,170;\h681;\m170,193;\D23;
\m246,193;\D23;\m322,193;\D23;\m397,193;\D23;\m473,193;\D23;\m549,193;\D23;\m624,193;\D23;\m700,193;\D23;\m776,193;\D23;\m851,193;
\D23;\m166,143;\s2,1;\s3,4;\D23;\m239,142;\U1;\s5,5;\h4;\s3,-3;\s1,-2;\D2;\s-1,-2;\s-2,-3;\s-11,-11;\h15;\m316,148;\h12;\s-6,-9;\h3;
\s2,-1;\s1,-1;\s1,-4;\D2;\s-1,-3;\s-2,-2;\s-3,-1;\r4;\s-3,1;\s-1,1;\s-1,2;\m400,148;\s-10,-16;\h16;\m400,148;\D23;\m478,148;\r10;
\s-2,-10;\s2,1;\s3,1;\h3;\s3,-1;\s2,-2;\s1,-4;\D2;\s-1,-3;\s-2,-2;\s-3,-1;\r3;\s-3,1;\s-2,1;\s-1,2;\m555,144;\s-1,2;\s-3,2;\r2;
\s-4,-2;\s-2,-3;\s-1,-5;\D6;\s1,-4;\s2,-2;\s4,-1;\h1;\s3,1;\s2,2;\s1,3;\U1;\s-1,4;\s-2,2;\s-3,1;\r1;\s-4,-1;\s-2,-2;\s-1,-4;
\m632,148;\s-11,-23;\m617,148;\h15;\m698,148;\s-3,-2;\s-2,-2;\D2;\s2,-2;\s2,-1;\s4,-1;\s3,-1;\s2,-2;\s1,-3;\D3;\s-1,-2;\s-1,-1;
\s-3,-1;\r4;\s-3,1;\s-2,1;\s-1,2;\U3;\s1,3;\s3,2;\s3,1;\s4,1;\s2,1;\s1,2;\U2;\s-1,2;\s-3,2;\r4;\m782,140;\s-1,-3;\s-2,-2;\s-3,-2;
\r1;\s-4,2;\s-2,2;\s-1,3;\U1;\s1,3;\s2,2;\s4,2;\h1;\s3,-2;\s2,-2;\s1,-4;\D5;\s-1,-6;\s-2,-3;\s-3,-1;\r3;\s-3,1;\s-1,2;\m836,143;
\s2,1;\s4,4;\D23;\m861,148;\s-3,-2;\s-2,-3;\s-2,-5;\D3;\s2,-6;\s2,-3;\s3,-1;\h2;\s3,1;\s3,3;\s1,6;\U3;\s-1,5;\s-3,3;\s-3,2;\r2;
\m851,170;\U681;\m829,170;\h22;\m840,204;\h11;\m840,238;\h11;\m840,272;\h11;\m840,306;\h11;\m829,341;\h22;\m840,375;\h11;\m840,409;
\h11;\m840,443;\h11;\m840,477;\h11;\m829,511;\h22;\m840,545;\h11;\m840,579;\h11;\m840,613;\h11;\m840,647;\h11;\m829,681;\h22;
\m840,715;\h11;\m840,749;\h11;\m840,783;\h11;\m840,817;\h11;\m829,851;\h22;\m170,851;\h681;\m170,829;\U22;\m246,829;\U22;\m322,829;
\U22;\m397,829;\U22;\m473,829;\U22;\m549,829;\U22;\m624,829;\U22;\m700,829;\U22;\m776,829;\U22;\m851,829;\U22;\m851,397;\s-25,-9;
\m851,414;\s-55,-20;\m851,432;\s-84,-31;\m851,449;\s-111,-41;\m851,466;\s-139,-50;\m851,484;\s-166,-61;\m851,501;\s-194,-70;
\m851,519;\s-221,-81;\m851,536;\s-248,-90;\m851,553;\s-274,-99;\m848,570;\s-297,-108;\m814,575;\s-288,-105;\m780,579;\s-279,-101;
\m746,584;\s-270,-98;\m711,589;\s-258,-94;\m678,595;\s-249,-91;\m645,600;\s-239,-87;\m613,606;\s-229,-84;\m581,612;\s-218,-80;
\m550,617;\s-209,-75;\m519,624;\s-199,-73;\m489,630;\s-188,-68;\m460,637;\s-177,-64;\m432,644;\s-168,-61;\m404,651;\s-158,-57;
\m377,659;\s-146,-53;\m351,667;\s-134,-49;\m326,675;\s-123,-45;\m303,684;\s-115,-41;\m280,693;\s-106,-38;\m258,703;\s-88,-32;
\m238,713;\s-68,-25;\m221,724;\s-51,-18;\m204,735;\s-34,-12;\m187,746;\s-17,-6;\m170,757;\s12,-8;\s13,-10;\s14,-9;\s13,-9;\s13,-7;
\s11,-6;\s24,-12;\s15,-6;\s17,-7;\s20,-7;\s42,-14;\s33,-9;\s29,-8;\s34,-8;\s13,-3;\s34,-8;\s42,-8;\s40,-8;\s35,-6;\s45,-8;\s31,-5;
\s51,-7;\s25,-4;\s75,-11;\m170,658;\s15,-15;\s15,-13;\s16,-14;\s14,-11;\s16,-11;\s12,-8;\s14,-9;\s18,-10;\s32,-17;\s20,-10;\s28,-12;
\s27,-12;\s31,-12;\s36,-13;\s9,-4;\s31,-10;\s45,-15;\s50,-15;\s25,-8;\s36,-9;\s40,-11;\s76,-20;\s33,-8;\s35,-7;\s7,-2;\m170,518;
\s15,-7;\s14,-6;\s16,-7;\s14,-5;\s14,-6;\s15,-5;\s14,-4;\s16,-4;\s22,-6;\s12,-3;\s13,-3;\s20,-4;\s42,-8;\s64,-13;\s12,-2;\s21,-3;
\s33,-5;\s22,-3;\s51,-7;\s24,-3;\s30,-4;\s38,-4;\s8,-1;\s76,-10;\s28,-3;\s30,-3;\s17,-1;\m170,518;\s12,-8;\s12,-7;\s12,-8;\s14,-7;
\s11,-6;\s13,-6;\s2,-1;\s20,-8;\s13,-4;\s18,-6;\s25,-7;\s19,-5;\s22,-6;\s24,-6;\s10,-2;\s22,-5;\s27,-5;\s27,-4;\s27,-5;\s41,-6;
\s40,-5;\s35,-5;\s8,-1;\s76,-11;\s35,-4;\s24,-3;\s17,-2;\s26,-2;\s24,-1;\s25,-2;\m170,509;\s12,-23;\s13,-25;\s14,-24;\s13,-24;
\s13,-21;\s11,-17;\s24,-36;\s15,-20;\s18,-23;\s19,-23;\s20,-26;\s20,-24;\s20,-23;\s15,-16;
\fi\end{picture}
\end{center}
\caption{Event rate versus signal-to-noise ratio ($S/N=1$ for
$E_{100}=55$~TeV) for the different  models of AGN neutrino production
(SDSS, Stecker et al. (1992); SPmax and SPmin, Szabo and Protheroe (1992);
Bi, Biermann (1992); SB, Sikora and Begelman (1992)).}
\end{figure}
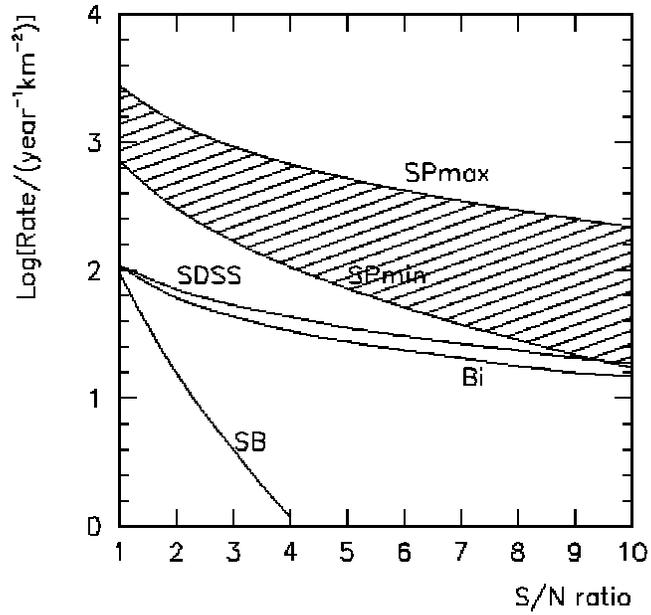


\begin{thebibliography}{99}
\bibitem{1}
G.A.~Askar'yan, Zh. Exp. Teor. Fiz. 41 (1961) 616; ibid. 48 (1965) 988.
\bibitem{2}
G.A.~Gusev and I.M.~Zheleznykh, Piz'ma Zh. Exp. Teor. Fiz. 38 (1983) 505;
M.A.~Markov  and I.M.~Zheleznykh, Nucl. Inst. Methods A248 (1986) 242.
\bibitem{3}
J.P.~Ralston and D.W.~McKay, Nucl. Phys. B (Proc. Suppl.) 14A (1990) 356.
\bibitem{4}
I.N.~Boldyrev et al., in: Proc. 3rd Int. Workshop on Neutrino Telescopes,
ed. Milla Baldo Ceolin (Venezia, 1991) p.~337.
\bibitem{5}
E.~Zas, F.~Halzen and T.~Stanev, Phys. Lett. B257 (1991) 432; Phys. Rev.
D45 (1992) 362.
\bibitem{6}
V.V.~Bogorodsky and V.P.~Gavrilo, Ice: Physical Properties. Modern Methods
of Glaciology (Leningrad, 1980).
\bibitem{7}
R.N.~Vostretsov et al., in: Data of Glaciological Studies (Moscow, 1984)
V.~51, p.~172.
\bibitem{8}
R.A.~Lawton and A.R.~Ondrejka, NBS Technical note 1008 (U.S. Department of
Commerce, 1978).
\bibitem{9}
C.E.~Cook and M.~Bernfeld, Radars Signals. An Introduction to Theory and
Application (New York -- London, 1967).
\bibitem{10}
A.F.~Bogomolov et al., in: Proc. 20th Int. Cosmic Ray Conf. (Moscow, 1987)
V.~6, p.~472.
\bibitem{11}
A.N.~Kalinovsky, N.V.~Mokhov and Yu.P.~Nikitin, Passage of High Energy
Particles through matter (Moscow, 1985).
\bibitem{12}
C.~Quigg, M.H.~Reno and T.P.~Walker, Phys. Rev. Lett. 57 (1986) 774.
\bibitem{13}
V.S.~Berezinsky, A.Z.~Gazizov, I.L.~Rozental, and G.T.~Zatsepin, Yad. Fiz.
43 (1986) 637.
\bibitem{14}
V.S.~Berezinsky and A.Z.~Gazizov, Yad. Fiz. 29 (1979) 1589.
\bibitem{15}
E.~Eichten, I.~Hinchliffe, K.~Lane and C.~Quigg, Rev. Mod. Phys. 56 (1984)
579.
\bibitem{16}
L.V.~Gribov, E.M.~Levin and M.G.~Ryskin, Phys. Rep. 100 (1983) 1.
\bibitem{17}
D.W.~McKay and J.P.~Ralston, Phys. Lett. B167 (1986) 103.
\bibitem{18}
A.M.~Dziewonski and D.L.~Anderson, Phys. Earth Planet. Inter. 25 (1981) 297.
\bibitem{19}
V.S.~Berezinsky and A.Z.~Gazizov, Piz'ma Zh. Exp. Teor. Fiz. 25 (1977) 276.
\bibitem{20}
I.M.~Zheleznykh and E.A.~Taynov, Yad. Fiz. 32 (1980) 468.
\bibitem{21}
V.S.~Berezinsky, in: Proc. Neutrino 77 Conf. (Moscow, 1977) p.~177;
V.S.~Berezinsky and V.L.~Ginzburg, Mon. Not. R. Astr. Soc. 194 (1981) 3.
\bibitem{22}
R.~Silberberg and M.M.~Shapiro, in: Proc. 16th Int. Cosmic Ray Conf.
(Kyoto, 1979) V.~10, p.~357
\bibitem{23}
F.W.~Stecker, C.~Done, M.H.~Salamon and P.~Sommers, Phys. Rev. Lett. 66
(1991) 2697
\bibitem{24}
F.W.~Stecker, C.~Done, M.H.~Salamon and P.~Sommers, in:  Proc. Workshop
High Energy Neutrino Astronomy, eds. V.J.~Stenger, J.G.~Learned, S.~Pakvasa
and X.~Tata (Singapure, 1992) p.~1.
\bibitem{25}
A.P.~Szabo and R.J.~Protheroe, ibid. p.~24.
\bibitem{26}
P.L.~Biermann, ibid. p.~86.
\bibitem{27}
M.~Sikora and M.C.~Begelman, ibid. p.~114.
\bibitem{28}
V.~Berezinsky, Phil. Trans. R. Soc. Lond. A346 (1994) 93.
\bibitem{29}
T.~Stanev, in: Proc. 23rd Int. Cosmic Ray Conf. (Calgary, 1993) Invited,
rapporteur and highlight papers, p.~503.
\bibitem{30}
L.V.~Volkova and G.T.~Zatsepin, Yad. Fiz. 37 (1983) 353; Izv. Acad. Nauk
SSSR (Fiz. Ser.) 49 (1985) 1386; L.V.~Volkova (1994), private
communication.
\bibitem{31}
J.G.~Learned and T.~Stanev, in: Proc. 3rd Int. Workshop on Neutrino
Telescopes, ed. Milla Baldo Ceolin (Venezia, 1991) p.~473.
\bibitem{32}
J.P.~Ralston, talk at the DPF Snowmass meeting (1994), unpublished.
\end{thebibliography}
\end{document}